\documentclass{emulateapj}
\usepackage{apjfonts}
\def\lya{Ly$\alpha$}
\def\lyb{Ly$\beta$}
\def\kms{km~s$^{\textrm{-}1}$}
\def\da{$D_{\!A}$}

\def\qi{\emph{q1}}
\def\qii{\emph{q2}}
\def\qiii{\emph{q3}}
\def\qiv{\emph{q4}}
\def\g6{\emph{g6}}
\def\vzw{\emph{wvzw}}
\def\nw{\emph{wnw}}
\def\cw{\emph{wcw}}

\slugcomment{}
\journalinfo{Published in The Astrophysical Journal (2008, \apj, 675, 946)}

\shorttitle{Correlation Anisotropies in the \lya\ Forest}

\shortauthors{Marble et al.}

\begin{document}
  
\title{The Flux Auto- and Cross-Correlation of the \lya\ Forest.
\\II. Modelling Anisotropies with Cosmological Hydrodynamic Simulations}

\author{Andrew R. Marble, Kristoffer A. Eriksen, Chris D. Impey, 
Benjamin D. Oppenheimer, Romeel Dav\'{e}}
\affil{Steward Observatory, University of Arizona, Tucson, AZ 85721}

\begin{abstract}

The isotropy of the \lya forest in real-space uniquely provides a
measurement of cosmic geometry at $z>2$.  The angular diameter
distance for which the correlation function along the line of sight
and in the transverse direction agree corresponds to the correct
cosmological model.  However, the \lya\ forest is observed in
redshift-space where distortions due to Hubble expansion, bulk flows,
and thermal broadening introduce anisotropy.  Similarly, a
spectrograph's line spread function affects the autocorrelation and
cross-correlation differently.  
In this the second paper of a series on using the \lya\ forest observed in 
pairs of QSOs 
for a new application of the Alcock-Paczy\'{n}ski (AP) test, these
anisotropies and related sources of potential systematic error are investigated with
cosmological hydrodynamic simulations.
Three prescriptions for galactic outflow were
compared and found to have only a marginal effect on the \lya\ flux
correlation (which changed by at most 7\% with use of the 
currently favored variable-momentum wind model vs. no winds at all).
An approximate solution for obtaining the zero-lag
cross-correlation corresponding to arbitrary spectral resolution
directly from the zero-lag cross-correlation computed at
full-resolution (good to within 2\% at the scales of interest) is
presented. Uncertainty in the observationally determined mean flux
decrement of the \lya\ forest was found to be the dominant source of
systematic error; however, this is reduced significantly when
considering correlation ratios.  We describe a simple scheme for
implementing our results, while mitigating systematic errors, in the
context of a future application of the AP test.

\end{abstract}

\keywords{cosmology: miscellaneous --- intergalactic medium ---
  methods: numerical --- quasars: absorption lines}

\section{Introduction}

Significant observational and theoretical advances in recent
decades have made the \lya\ forest a powerful and unique cosmological tool
for studying the high-redshift universe.
Originally named \citep{1981wey19araa41} for the dense pattern of
seemingly discrete \lya\ absorption lines 
seen in high-redshift QSO spectra \citep{1971lyn164apjl73},
the absorption is now understood to 
trace a continuous distribution of non-uniform neutral hydrogen gas 
that in turn maps the underlying dark matter 
(see \citealt{1998rau36araa267} for a review).  The competing
processes of recombination and photoionization lead to a tight
relationship between the density of the gas and the neutral fraction,
giving rise to a relatively straightforward link between \lya\
absorption and the large scale structure of the universe.  
Cosmological
simulations employing this prescription have had remarkable success
reproducing detailed properties of the \lya\ forest provided by 
high-resolution ground-based QSO spectra 
\citep{1994cen437apjl9, 1995zha453apjl57, 1996her457apjl51,
  1998the301mnras478} and low-$z$ \emph{HST} observations
\citep{1995pet295aap9, 1999dav511apj521}, paving the way for the \lya\
forest to be reliably used for cosmological investigation.

\citet{1999hui511apjl5} and \citet{1999mcd518apj24} first suggested
using autocorrelation and cross-correlation measurements in the \lya\ forest
for a new application of the Alcock-Paczy\'{n}ski (AP) test 
\citep{1979alc281nat358}, a purely geometric method for measuring
cosmological parameters that is primarily sensitive to
$\Omega_{\Lambda}$ at $z>1$.  The essence of this cosmological test is
that spherical objects observed at high redshift will only appear to
be equal in their radial and transverse extent if the correct
angular diameter distance is used 
to determine the latter.  More generally, the correlation function of
an isotropic medium, such as the \lya\ forest, 
measured as a function of separation 
along the line of sight (the autocorrelation
$\xi_\parallel$) and in the transverse direction
(the cross-correlation $\xi_\perp$) will agree only if the
correct cosmology is assumed.

Spectroscopy of the \lya\ forest in any single QSO spectrum yields the
complete autocorrelation, albeit with significant variance from one line of
sight to another.  The cross-correlation, on the other hand, must be
pieced together from pairs of QSOs with different transverse separations.
Until recently, only approximately a dozen pairs 
with similar redshifts (so that their \lya\ forests overlap)
and separations of a few arcminutes or less (the
correlation signal diminishes rapidly beyond this point)
were known (see, \emph{e.g.}, \citealt{2003rol341mnras1279} and
references therein).  
The 2dF QSO Redshift Survey \citep[2QZ;][]{2004cro349mnras1397}
significantly increased this number, and with motivation from
\citet{2003mcd585apj34}, moderate
resolution spectra (FWHM $\simeq$ 2.5 \AA) with modest signal-to-noise
ratios (S/N $>$ 10 per pixel) have been obtained for more than 50 of these
pairs at the VLT \citep{2006cop370mnras1804}, MMT, and Magellan
\citep[][hereafter referred to as Paper I]{2007mar} observatories.

While conceptually simple, the \lya\ forest variant of the AP test is
not as straightforward as measuring the autocorrelation and
cross-correlation from pairs of QSOs and determining the angular
diameter 
distance which satisfies the presumption of isotropy.  Rather, two
additional sources of anisotropy must be accounted for.  First, the
line spread function (LSF) of the spectrograph smooths QSO spectra along the
line of sight, affecting the autocorrelation and cross-correlation
differently.  Second, nonzero velocities caused by the expansion of
the universe, gravitational collapse, and thermal broadening make the
correlation function in redshift-space ($z$-space) anisotropic
\citep{1987kai227mnras1}. Fortunately, the theoretical work of
\citet{1999hui511apjl5}, \citet{1999mcd518apj24}, and
\citet{2003mcd585apj34} found that these redshift-space distortions
can be disentangled from the desired cosmological signature.  

The aim of this paper is to investigate these non-cosmological
anisotropies in the \lya\ flux correlation function in a manner which
is directly applicable to observations of QSO
pairs that are suitable for a new application of the AP test.  To this
end we have employed a variety of cosmological hydrodynamic
simulations to model the autocorrelation and cross-correlation of the
\lya\ forest in both redshift-space and real-space.  These simulations
are described in \S~\ref{sec_simdata}, as well as 
our procedure for extracting mock \lya\ absorption spectra from
them.  In \S~\ref{sec_corrfunc} and
\S~\ref{sec_hybrid} we introduce the correlation function and 
discuss how to mitigate relevant size and mass resolution limitations
of the current generation of simulations.  The effect of arbitrary
spectral resolution on the correlation function is the subject of
\S~\ref{sec_resolution}.
Additional potential sources of systematic error, both computational
and observational, are addressed in
\S~\ref{sec_systematics}.  The implications of our results for the AP
test are the topic of \S~\ref{sec_ap}.  Finally, in
\S~\ref{sec_summary}, we summarize this work and its findings.

\section{Simulation Data}\label{sec_simdata}

\subsection{Simulations}\label{sec_sims}

This body of work draws from a suite of eight cosmological
simulations which primarily differ in their
size, mass resolution, and
prescription for galactic outflow (Table~\ref{tab_sims}).  
Together, \emph{w16n256vzw} (abbreviated as \vzw) and \g6\ mitigate the effects of limitations in
volume and mass resolution as discussed in \S~\ref{sec_hybrid}.
Differing wind models (described in \S~\ref{sec_wind})
are investigated with the \emph{w16n256cw} and
\emph{w16n256nw} simulations (abbreviated as \cw\ and \nw\ respectively), which are 
otherwise identical to \vzw.  Similarly, the \qi-\qiv\ simulations
differ only by their number of particles, $N_p$, and are used in \S~\ref{sec_hybrid} to test
for convergence as a function of mass resolution.  
All of these simulations have been the subject of previous
study; therefore, we address only the relevant details and 
direct interested readers to the references provided for additional
discussion.  The common genesis of these simulations is described below,
while their differences are contrasted in Table~\ref{tab_sims} and the sections referenced above.

The $N$-body $+$ hydrodynamic code {\sc Gadget} \citep{2001spr6na79}, with
modifications described in \citet{2003spr339mnras312}, was
used to create \qi, \qii, \qiii, \qiv, and \g6 \citep[][but also see
\citet{2006fin639apj672} regarding \g6]{2003spr339mnras312},
while \vzw, \cw, and \nw\ \citep{2006opp373mnras1265}
were run with a similarly modified version of its successor 
{\sc Gadget-2} \citep{2005spr364mnras1105}.  This
code computes gravitational forces via a tree particle-mesh solver 
and hydrodynamical forces with an entropy-conservative
formulation of smoothed particle hydrodynamics (SPH).  Modelling of
additional physical processes includes prescriptions
for star formation and supernova feedback within evolving galaxies, which
impact the intergalactic medium (IGM) via outflow from galactic winds.
A spatially uniform photoionization background is included, with the
spectral shape and redshift evolution given by
\citet{1996haa461apj20} and \citet{2001haa64cghr} for the {\sc Gadget} and
{\sc Gadget-2} runs respectively.  Radiative heating and cooling is
calculated assuming photoionization equilibrium and optically thin
gas.
All of the simulations were run as cubic volumes with periodic boundary
conditions and the same cosmological parameters 
($\Omega_m=0.3$, $\Omega_\Lambda=0.7$, $\Omega_b=0.04$, $\sigma_8=0.9$, and
$h=0.7$), which we assume throughout this paper.

\subsection{Line of Sight Selection}

``Observed'' lines of sight through the simulation box (parallel to the three
principal axes) were grouped into sets, each of which probed the desired separation
range ($0-5$ arcminutes). $N_s$ sets were randomly distributed across
each of the three mutually orthogonal faces of the box in order to 
representatively sample the diversity of structure present in the
simulation volume. The value of $N_s$ (Table~\ref{tab_sims}),
which roughly scales inversely with the box length of the simulation cube
for comparable total path length, yields oversampled structure (correlated 
measurements) in some cases.  However, there are sufficient independent lines of sight
through each simulation to achieve negligible uncertainty in the \lya\ forest flux 
correlation mean despite considerable variance.

The lines of sight in a given set were arranged in the following
manner.  A position on the face of the
simulation box was randomly selected, from which an imaginary $300\arcsec$
long line was extended at a random angle within the same plane.  If the
line happened to intersect an edge of the simulation face, it was
continued on the opposite side, per the wrapped boundary conditions.
The two opposing ends of the line and 11 intermediate positions
(2, 3, 8, 12, 15, 25, 45, 80, 150,
210, and 250 arcseconds from the origin) defined starting coordinates for
that set.  The intervals between these 13 lines of sight, determined
via a Monte Carlo approach designed to maximize sampling of angular
separation (particularly at small separations where the correlation
function evolves more rapidly) with a minimal number of spectra, yield
73 pairings with unique separations.

\subsection{\lya\ Flux Spectra}\label{sec_spectra}

Two \lya\ transmitted flux spectra, one corresponding to redshift-space and another to 
real-space, 
were computed as a function of position, $x$, along each line of sight using a modified 
version of
the program {\tt specexbin} \citep[originally part of {\tt tipsy}; ][]{1999dav511apj521}.
First, the physical properties of the gas (density, temperature, and velocity) were 
calculated at $\simeq20~h^{\textrm{-}1}$ comoving kpc intervals
($\Delta v\,\simeq\,2.1\,$\kms\ or $\Delta \lambda\,\simeq\,0.029\,$\AA\ at $z=2.4$).  
For the real-space 
spectra, the velocities (resulting from Hubble expansion across the length of the box, bulk flows, and thermal 
broadening) were reset to zero.  Then, the corresponding \ion{H}{1} opacities, 
$\tau$, were determined, using ionization fraction lookup tables generated with {\tt Cloudy v96} \citep{1998fer110pasp761}, and converted to \lya\ transmitted flux,

\begin{equation}\label{eq_flux}
f(x) = e^{-\tau(x)}.
\end{equation}

In an additional intermediate step, the extracted opacities were multiplied by a single scaling factor in
order to match the mean transmitted flux, $\langle f \rangle$, in redshift-space to the observed value of
either \citet{1993pre414apj64} or \citet{2005kir360mnras1373}. 
For the moderately overdense regions characteristic of the \lya\ forest
($\rho/\bar{\rho}<10$), this has the same effect as changing the
amplitude of the photoionizing background (which determines $\langle f
\rangle$)
when running the simulation and/or when later computing the ionization fraction
to determine opacity \citep{1998cro495apj44}.
The appropriate scaling was determined for each simulation and redshift
via an iterative process. The raw opacity
values from all redshift-space extractions were multiplied by a
single scaling factor (originally one),
converted to fluxes, and averaged.  The scaling factor was then adjusted to be 
higher or lower as needed, and these steps were repeated with
incrementally smaller adjustments until the resulting mean flux
agreed with the desired value.
This convergence was considered complete
when the difference was less than or equal to the formal error in the
calculated mean.  Given the large number of extracted opacities, this
typically corresponded to a relative difference of less than 0.01\%.

\begin{figure} 
\begin{center}
\includegraphics[width=3.5in]{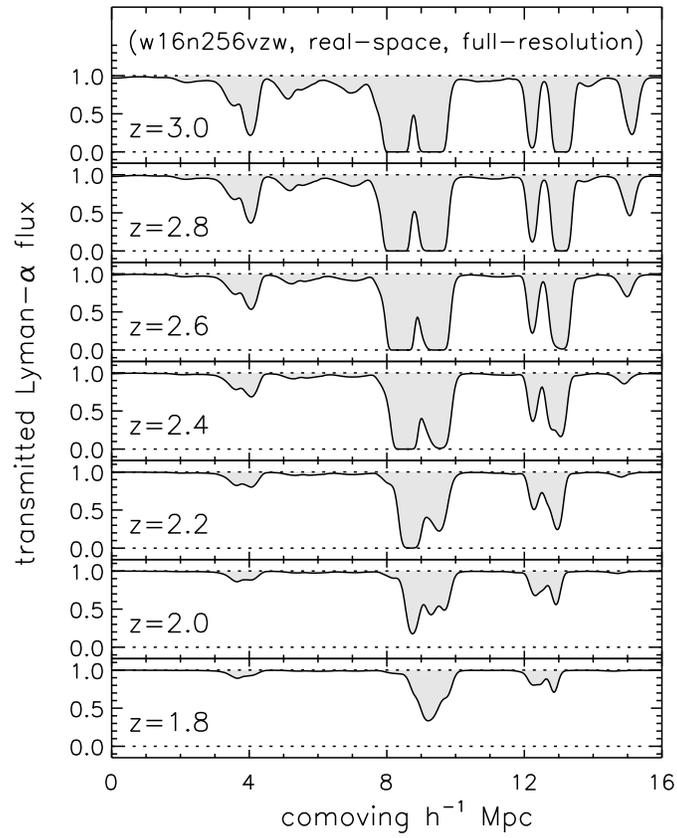}
\caption{Redshift evolution of the \lya\ forest as seen in the same
  line of sight at $1.8<z<3$.}\label{fig_spec_z}
\end{center}
\end{figure}

\begin{figure} 
\begin{center}
\includegraphics[width=3.5in]{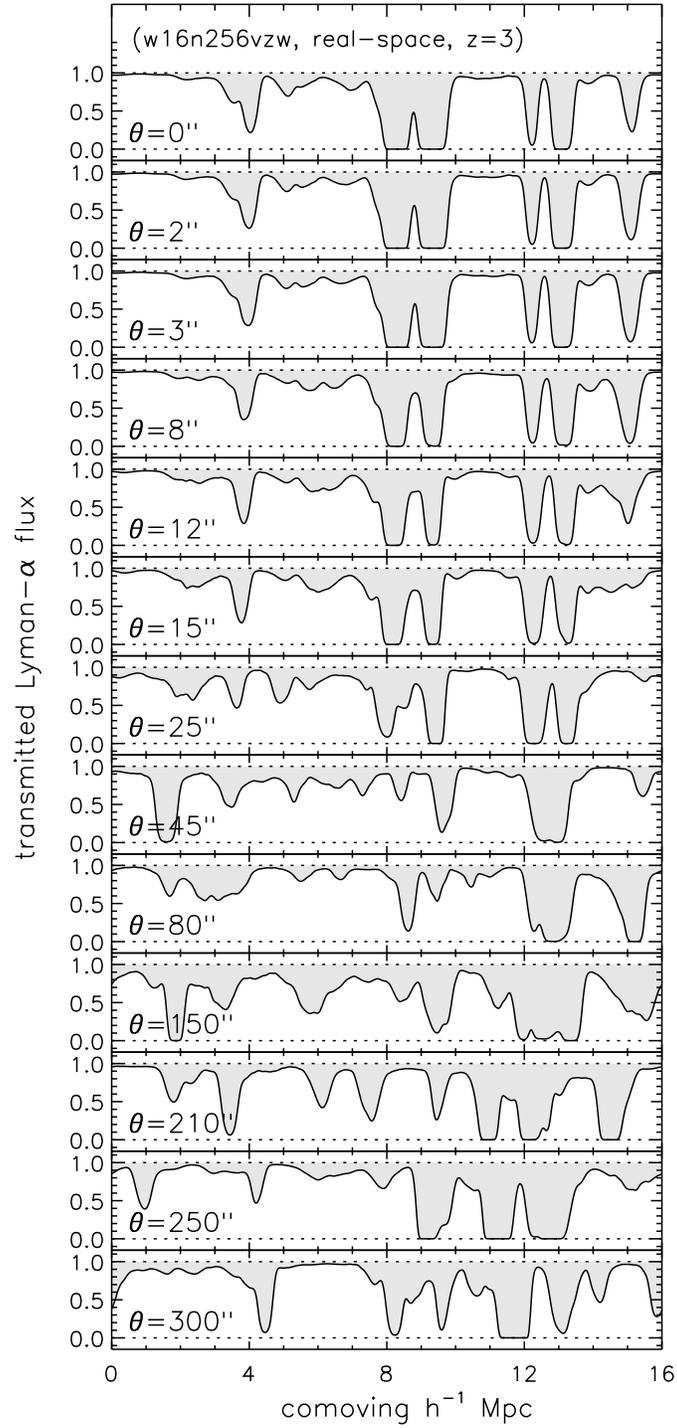}
\caption{One set of lines of sight illustrating how the cross-correlation diminishes
  with increasing separation.}\label{fig_spec_sep}
\end{center}
\end{figure}

\begin{figure} 
\begin{center}
\includegraphics[width=7.25in]{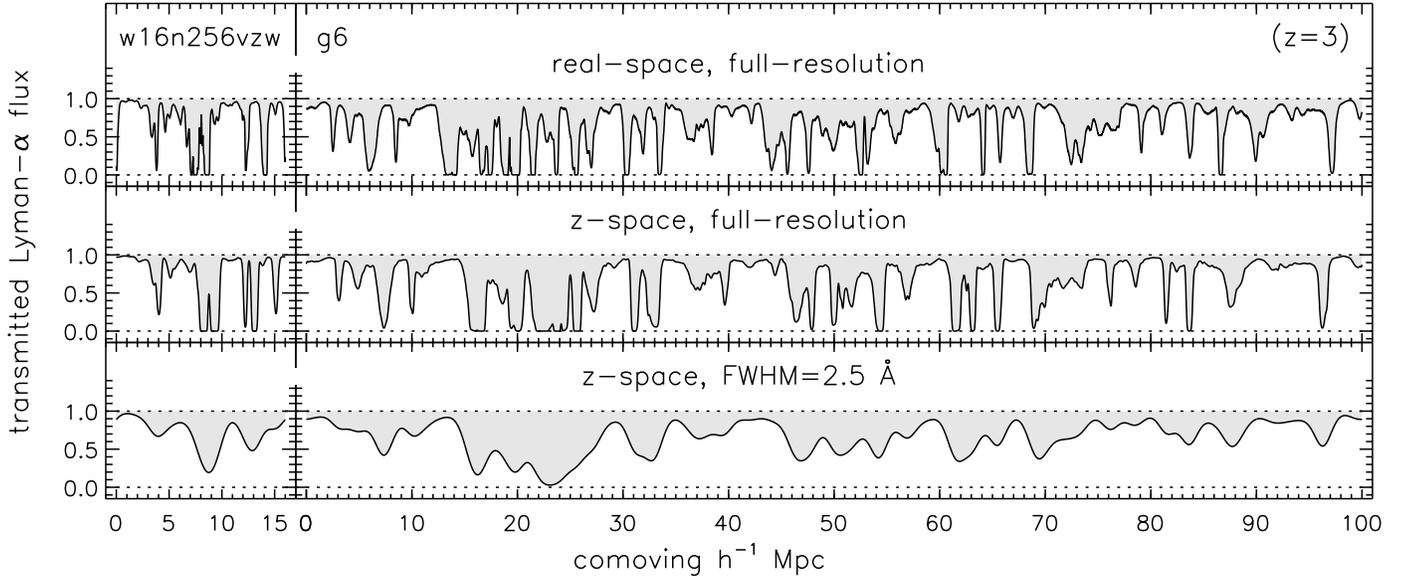}
\end{center}
\caption{Simulated spectra from \vzw\ and \g6\ show the larger box
 length (path length) of the latter, the higher mass resolution
 (narrower features) of the former, the subtle redistribution of
 opacity due to redshift-space distortions, and the smoothing effect of
 spectral resolution.}\label{fig_spec_other}
\end{figure}

Figure~\ref{fig_spec_z} shows the resulting spectrum for a single line
of sight through the \vzw\ simulation box at different redshifts. The
corresponding set of lines of sight (for $z=3$) is shown in
Figure~\ref{fig_spec_sep}, illustrating the
decreasing correlation (in the form of visual similarity)
at increasingly larger transverse separations.
A comparison of the same line of sight in real-space and
redshift-space is provided in Figure~\ref{fig_spec_other}, where the
redistribution of opacity due to redshift-space distortions is subtle,
but evident.  In addition, narrower absorption features can be seen relative
to a different line of sight through the larger \g6\ simulation, due to the
poorer mass resolution of the latter.

\section{The Correlation Function}\label{sec_corrfunc}

From the ensemble of lines of sight through each simulation, we know
the transmitted \lya\ flux as a function of velocity, 

\begin{equation}\label{eq_v}
v \equiv \frac{H(z_{sim})\,x}{1+z_{sim}},
\end{equation}

\noindent in the radial ($v_\parallel$) and
transverse ($v_\perp$) directions for $\Sigma n = N_s$ different realizations
(\emph{i.e.,} sets of lines of sight).
Here $H(z_{sim})$ is the value of the Hubble parameter at the fixed redshift of the simulation,
and the denominator accounts for $x$ being in comoving coordinates.
For notational convenience, we define $\delta$ to be the relative difference between
$f$ and the global mean, 

\begin{equation}\label{eq_bigf}
  \delta_n\left(v_\parallel, v_\perp\right) \equiv \frac{f_n\left(v_\parallel, v_\perp\right)}{\langle f
  \rangle} - 1.
\end{equation}

\begin{figure} 
\begin{center}
\includegraphics[width=3.5in]{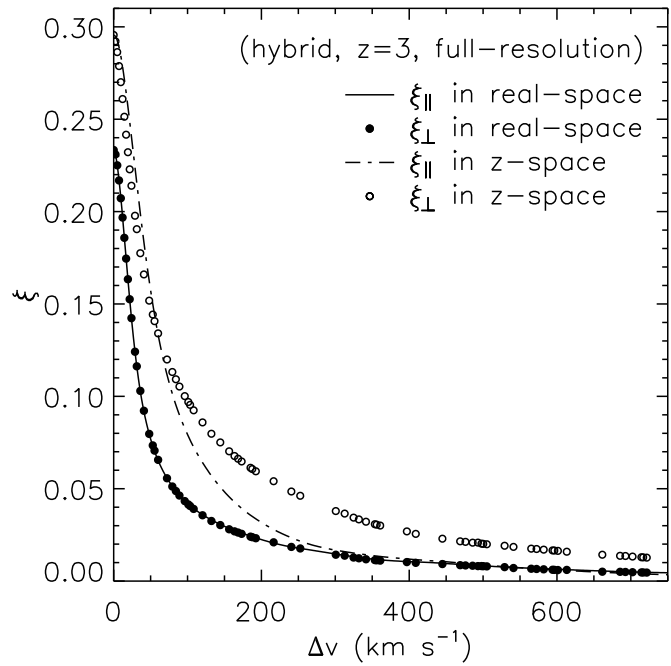}
\end{center}
\caption{In real-space, the correlation function is isotropic.
  However, in redshift-space, distortions caused by line-of-sight
  velocities affect the autocorrelation ($\xi_\parallel$) and
  cross-correlation ($\xi_\perp$) differently.}\label{fig_zspace}
\end{figure}

\noindent The relation between transmitted flux separated
along the line of sight or in the transverse direction 
by a velocity difference $\Delta v$ 
is
given by the autocorrelation, 

\begin{equation}\label{eq_auto}
  \xi_{\parallel}(\Delta v) = 
  \frac{1}{N_s} \sum_{n=1}^{N_s} 
  \frac{1}{N_\perp} \sum_{i=1}^{N_\perp}
  \frac{1}{N_\parallel} \sum_{j=1}^{N_\parallel} 
  \delta_n\left(v_{\parallel_j}, v_{\perp_i}\right) \,
  \delta_n\left(v_{\parallel_j}+\Delta v, v_{\perp_i}\right),
\end{equation}

\noindent and zero-lag cross-correlation, 

\begin{equation}
  \xi_{\perp}\!\left(\Delta v\right) = 
  \frac{1}{N_s} \sum_{n=1}^{N_s} 
  \frac{1}{N_\parallel} \sum_{j=1}^{N_\parallel} 
  \delta_n\left(v_{\parallel_j}, v_{\perp_i}\right) \, 
  \delta_n\left(v_{\parallel_j}, v_{\perp_i}+\Delta v\right),
\end{equation}

\noindent respectively.  Note that $f_n$ (and therefore $\delta_n$) is periodic due to
the wrapped boundary conditions of the simulation box.
In real-space, the correlation of the \lya\ forest is isotropic, and
$\xi_{\parallel} = \xi_{\perp}$.  Figure~\ref{fig_zspace} confirms
this basic result for the hybrid correlation measurements (discussed
in \S~\ref{sec_hybrid}) at $z=3$ and shows the
anisotropy introduced in redshift-space.

\subsection{Box Length \emph{vs.} Mass Resolution}\label{sec_hybrid}

The \lya\ forest is believed to have formed, via gravitational
collapse, from perturbations in the initial density field.  
In order to reliably model the correlation function of the \lya\
forest, simulations must evolve a sufficiently large volume with adequate mass
resolution.  A simulation box length that is too small, or a gas particle
mass that is too large, excludes relevant perturbations on large and
small scales respectively.
In the moderately overdense regime of the \lya\ forest, growth is
sufficiently non-linear that perturbations of different sizes
become coupled, and the correlation function is affected even at scales
not excluded.
In addition, aliasing 
due to the periodic boundary conditions of the simulations
is extended from half the box length ($L$) to smaller scales
(we limit our analysis to
separations less than $L/3-L/4$).
Since simulations which can satisfy both of these
competing demands are not yet available, we mitigate these effects
by forming hybrid correlation curves from two
different simulations which meet the requirements independently.

The \vzw\ simulation has $256^3$ gas particles within an $L=16~h^{\textrm{-}1}$
Mpc box, yielding a gas particle mass of $m_{gas}=2.71 \times 10^6~h^{\textrm{-}1}
~M_{\sun}$.  In order to verify that this mass resolution is
sufficient for our purposes, we used the \qi, \qii, \qiii, and \qiv\
simulations to test for convergence (note that the result may be
simulation code dependent).  The \emph{q}-series 
are identical except for particle number, with gas 
particle masses which decrease with
increasing series number (42.4, 12.5, 3.72,
and 1.10 in units of $10^6~h^{\textrm{-}1} ~M_{\sun}$).  
As a consequence of their small box length ($L=10~h^{\textrm{-}1}$ Mpc), the
correlation is artificially depressed and the autocorrelation crosses
zero on the scales of interest to us.  Therefore, in order to make
meaningful comparisons (avoiding division by zero), the \emph{q}-series correlation curves were
all increased by an equal, constant amount such that \qi\ agrees with
\g6\ (which has a comparable gas mass resolution, but a much larger box length) at 445 \kms\ (this velocity choice is motivated below).

\begin{figure} 
\begin{center}
\includegraphics[width=7.25in]{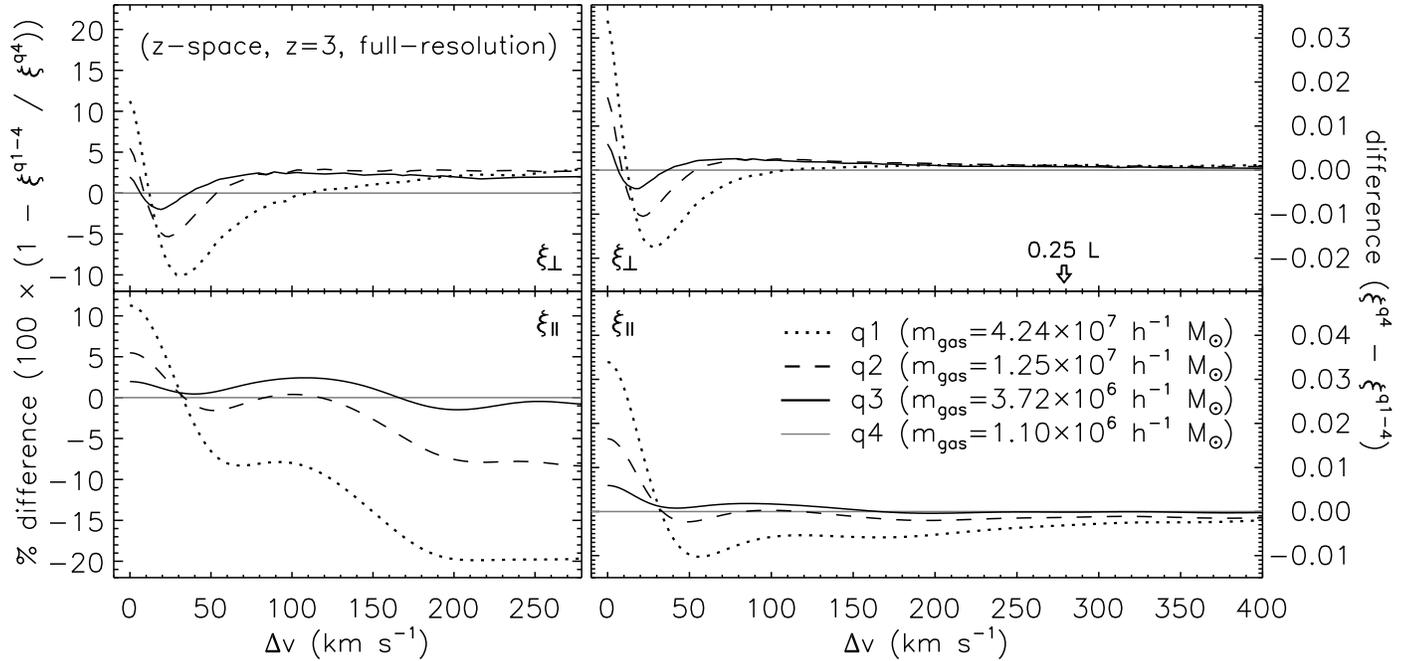}
\end{center}
\caption{Convergence of the cross-correlation (\emph{top}) and
  autocorrelation (\emph{bottom}) as a function of gas mass
  resolution for the \emph{q}-series simulations
  ($L=10~h^{\textrm{-}1}$ comoving Mpc)
  which differ only by the number of particles.}\label{fig_qseries}
\end{figure}

As shown in the left panels of Figure~\ref{fig_qseries}, the
cross-correlation (\emph{top}) and autocorrelation (\emph{bottom}) from \qiii\
agree well with \qiv\ ($<3\%$ relative difference for $\Delta
v$ corresponding to less than $L/4$).  We conclude that \vzw, which has a smaller gas particle
mass than \qiii, is not significantly compromised by mass
resolution. However, in addition to missing large scale power due to the $L=16~h^{\textrm{-}1}$ Mpc box length, the reliable separation range ($\scriptstyle\lesssim$ $L/4$) probed by \vzw\ corresponds to only $394-444$ \kms\ at $z=2-3$.
Conversely, the box length of the larger ($L=100~h^{\textrm{-}1}$ Mpc), 
but much lower-resolution ($484^3$ gas particles, $m_{gas}=9.79\times10^7~h^{\textrm{-}1} ~M_{\sun}$),
\g6\ simulation should more than suffice.
\citet{2003mcd585apj34} found little difference in the correlation
function between $L=40$ and $80~h^{\textrm{-}1}$ Mpc simulations.

\begin{figure} 
\begin{center}
\includegraphics[width=3.5in]{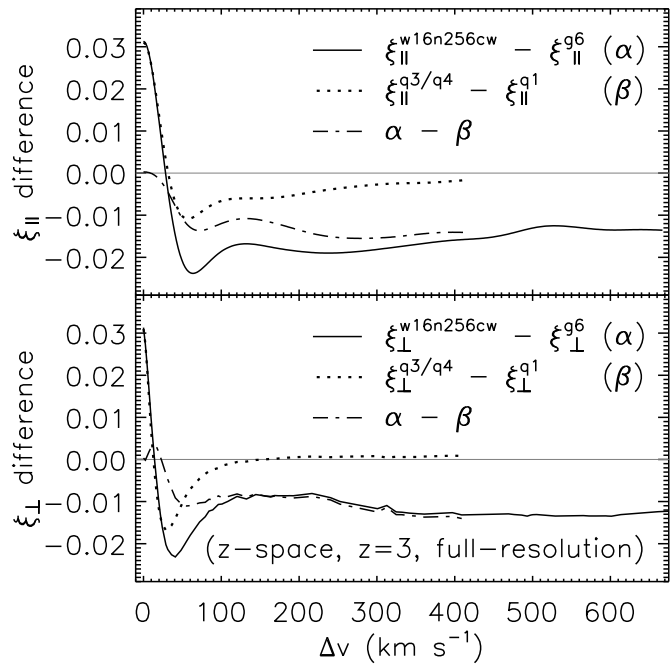}
\end{center}
\caption{Correlation difference between \g6\ and \cw\ (\emph{solid lines})
  primarily reflects the insufficient mass resolution of the former
  and box length of the latter.
  The signature of the mass resolution effect is illustrated by the
  dotted line, although for a smaller mass resolution difference.
  Accounting for the
  trend shown in Figure~\ref{fig_qseries}, the residual boxsize effect
  (\emph{dash-dotted lines}) should be significantly flatter, particularly on the
  smallest scales ($\Delta v\,\scriptstyle\lesssim$ 50 \kms).}\label{fig_boxsize}
\end{figure}

In Figure~\ref{fig_boxsize}, we consider the subtracted difference between the
correlation functions of \cw\ and \g6\ (\emph{solid lines}) in order to 
characterize the effects of insufficient simulation volume
and mass resolution
and to motivate a methodology for forming hybrid correlation
curves that mitigate them.  Note that \cw\ is used in lieu of \vzw\
in order to elliminate any additional differences due to wind models.
The signature of poor mass resolution is illustrated (\emph{dotted lines} in 
Fig.~\ref{fig_boxsize}) by the correlation difference
between \qi\ and the mean of \qiii\ and \qiv\ (which closely corresponds to the
mass resolution of \cw).  
The nature of the suppression of \cw\ due to its small
box length is then reflected in the residual (dash-dotted lines in 
Fig.~\ref{fig_boxsize}) between
these two curves.  However, since the mass resolution of \qi\ is
superior to \g6\ by more than a factor of 2, 
the dotted line underestimates the effect for \g6.  Extrapolation from
the right panels of Figure~\ref{fig_qseries} is
poorly constrained, although accounting for the trend implies significant
flattening of the dash-dotted line (Figure~\ref{fig_boxsize}) on small
scales.  Such a relatively
smooth alteration of the correlation function due to insufficient box
length is consistent with the expectation of constant suppression
when the evolution of coupled modes is not accounted for
\citep[][see Figure 15]{2000mcd543apj1}.  Although the true effect is
likely not a constant offset at all scales, this appears to be a
reasonable approximation.  Note also that the disparity
between \cw\ and \g6\ at sufficiently large $\Delta v$ appears to be
purely a box length effect.
Therefore, we define the hybrid correlation function $\xi^h$ to be equal to $\xi^{g6}$ for $\Delta v$ greater than an
adopted splice velocity, $v_s$. For $\Delta v < v_s$, $\xi_{\parallel}^h$ is equal
to $\xi_{\parallel}^{wvzw}$ plus the difference between $\xi_{\parallel}^{g6}$ and
$\xi_{\parallel}^{wvzw}$ at the splice velocity.  In the case of the
cross-correlation, this is slightly modified to preserve the boundary
condition $\xi_{\perp}\left(0\right)=\xi_{\parallel}\left(0\right)$.  More explicitly,

\begin{equation}
\label{eq_hybrid}
\begin{array}{rcl}

\xi^h\left(\Delta v > v_s\right) & = & \xi^{g6}\left(\Delta v\right)\\

\xi_\parallel^h\left(\Delta v < v_s\right) & = &
\xi_\parallel^{wvzw}\left(\Delta v\right) + \tilde{\delta}_\parallel\\

\xi_\perp^h\left(\Delta v < v_s\right) & = &
\xi_\perp^{wvzw}\left(\Delta v\right) + \tilde{\delta}_\perp +
\left(\frac{\delta_\perp-\delta_\parallel}{v_s}\right) \Delta v\\

\tilde{\delta} & \equiv & \xi^{g6}\left(v_s\right) - \xi^{wvzw}\left(v_s\right).

\end{array}
\end{equation}

\noindent The value of $v_s$ was chosen to be the minimum velocity
at which the effect of the \g6\ mass resolution can be assumed to be
negligible.  In order to ensure isotropy in real-space, $v_s$ must be
the same for the autocorrelation and cross-correlation.
Thus, based on Figures~\ref{fig_qseries}~and~\ref{fig_boxsize}, 
445 \kms\ was adopted as the splice velocity (for all redshifts).
It is worth noting that the resulting hybrid correlation curves are insensitive to the
exact choice of $v_s$ due to the relative flatness of the difference
between $\xi^{g6}$ and $\xi^{wvzw}$ on these scales.

\subsection{Accounting For Spectral Resolution}\label{sec_resolution}

\begin{figure} 
\begin{center}
\includegraphics[width=3.5in]{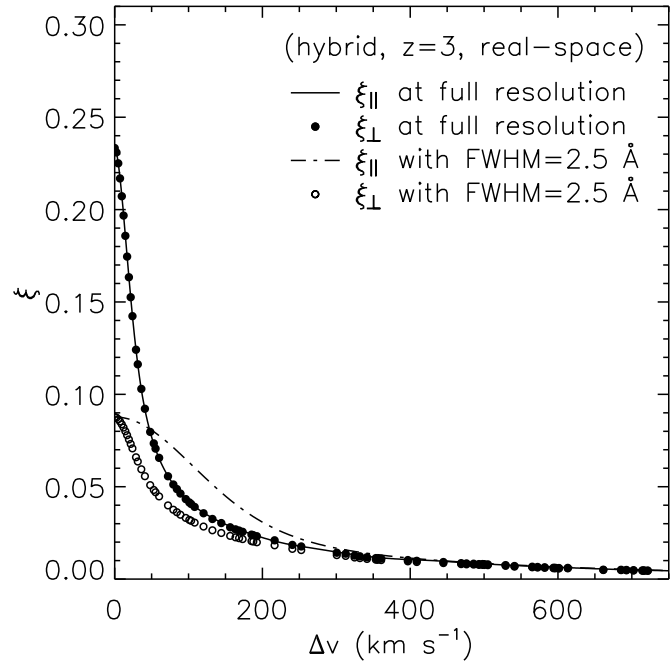}
\end{center}
\caption{Spectral smoothing redistributes autocorrelation ($\xi_\parallel$)
  while suppressing cross-correlation ($\xi_\perp$) on all scales, introducing anisotropy to the
  correlation function.}\label{fig_res}
\end{figure}

A real spectrum (\emph{i.e.}, observed with a telescope) is a
convolution, $\mathcal{S_\parallel}$, of the true 
transmitted flux along the line of sight with the LSF of the
spectrograph.  The LSF is generally Gaussian, and the width, $\sigma$, 
determines the resolution of the data,  

\begin{equation}\label{eq_smoothing} 
  \mathcal{S_\parallel}\left[\tilde{f}\left(v_{\parallel_j}\right), \sigma\right] 
  \equiv \sum_{k=j-\alpha}^{j+\alpha} \tilde{f}\left(v_{\parallel_k}\right) \,
  \frac{1}{\sqrt{2\pi}\,\sigma} \,\, e^{\frac{-(v_{\parallel_j}-v_{\parallel_k})^2}{2\,\sigma^2}},
\end{equation}

\noindent where $\alpha$ must be sufficiently large with respect to
$\sigma$ that the tails of
the exponential are effectively zero at the limits of convolution.
Figure~\ref{fig_spec_other} provides a comparison of a
simulated spectrum at full-resolution and the same spectrum degraded to
FWHM $= 2 \sqrt{2 \ln 2}$ and $\sigma =\ 2.5\,$\AA.
Similar to the anistropy introduced by redshift-space
distortions, this smoothing along the line of sight changes the
autocorrelation differently than the cross-correlation 
(Figure~\ref{fig_res}).  Thus, the latter anisotropy must be properly
accounted for in order to correct the former.

A sensible way of determining $\xi^\sigma$, the correlation function
corresponding to data of resolution 
$\sigma$, is to smooth the simulated spectra and then 
compute their correlation,

\begin{eqnarray}
  \xi^\sigma_\parallel\left(\Delta v\right) & \equiv &
  \frac{1}{N_s} \sum^{N_s}_{n=1} 
  \frac{1}{N_\perp} \sum^{N_\perp}_{i=1} 
  \frac{1}{N_\parallel} \sum^{N_\parallel}_{j=1}
  \mathcal{S_\parallel}\left[\delta_n\left(v_{\parallel_j}, v_{\perp_i}\right),
  \sigma\right] \nonumber \\  
  & & \times \, \mathcal{S_\parallel}\left[\delta_n\left(v_{\parallel_j}+\Delta v, v_{\perp_i}\right),
  \sigma\right]\label{eq_autosigma1} 
\\
  \xi^\sigma_\perp\left(\Delta v\right) & \equiv &
  \frac{1}{N_s} \sum^{N_s}_{n=1} 
  \frac{1}{N_\parallel} \sum^{N_\parallel}_{j=1}
  \mathcal{S_\parallel}\left[\delta_n\left(v_{\parallel_j}, v_{\perp_i}\right),
  \sigma\right] \nonumber \\ 
  & & \times \, \mathcal{S_\parallel}\left[\delta_n\left(v_{\parallel_j}, v_{\perp_i}+\Delta v\right),
  \sigma\right]\label{eq_crosssigma1}. 
\end{eqnarray}

\noindent This, however, has several disadvantages.
Individually smoothing each spectrum is a time-consuming
process which must be repeated for each desired value of
$\sigma$.  Likewise, the correlation calculations must be duplicated, 
and future consideration of different resolutions requires
the original simulated spectra.  More importantly, 
the creation of hybrid \emph{auto}correlation
curves as described in \S~\ref{sec_hybrid} 
is only valid if performed at full-resolution.  This is because spectral
smoothing redistributes correlation along the line of sight, negating
the validity of the splice point.
Finally, smoothing the spectra also
redistributes aliasing effects (which are mitigated in the hybrid
correlation function) to smaller separations, 
limiting the scales which can be reliably probed at a given resolution.
For the \vzw\ simulation, $\pm 3\sigma$ (where the LSF becomes negligible)
corresponds to a third of the
box length for a FWHM of 1.8/2.7 \AA\ at $z=2/3$.

An alternative method of accounting for spectral resolution, 
applied directly to the full-resolution hybrid correlation function,
solves each of these problems.
Convolving the full-resolution autocorrelation function with 
a Gaussian LSF of width $\sqrt{2}\,\sigma$ is mathematically identical
to recalculating the autocorrelation with spectra smoothed by a
Gaussian LSF of width $\sigma$,

\begin{equation}\label{eq_autosigma2} 
\mathcal{S_\parallel}\left[\xi_{\parallel}\left(\Delta
v\right), \sqrt{2} \sigma\right] = 
\xi^\sigma_\parallel\left(\Delta v\right).
\end{equation}

\noindent This convenient result is due to the fact that 
the convolution of two Gaussians is itself a Gaussian and that the 
autocorrelation and spectral smoothing 
are both a function of radial velocity (see the Appendix).

\begin{figure} 
\begin{center}
\includegraphics[width=3.5in]{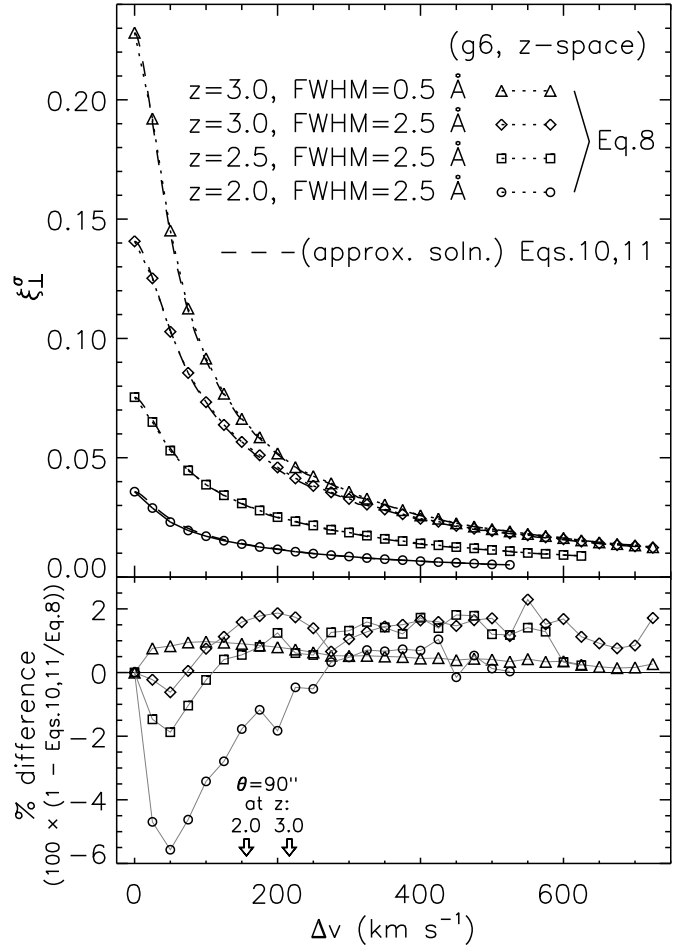}
\end{center}
\caption{Approximate solution for the suppression of
  cross-correlation due to spectral resolution,
  $\xi^\sigma_\perp \approx \xi_\perp
  \,
  \left(1+\left(\frac{1}{\xi^\sigma_\parallel\left(0\right)}-\frac{1}{\xi_\perp\left(0\right)}\right)
  \, \xi_\perp\right)^{\textrm{-}1}$, 
  differs from the result
  based on smoothed spectra by $\scriptstyle\lesssim$ 2\% for separations greater
  than 90 arcseconds.}\label{fig_res_cross}
\end{figure}

Unfortunately, this is not the case for the cross-correlation, and
there is no
corresponding analytical expression.  However, since
spectral smoothing redistributes correlation along the line of sight,
its effect in the orthogonal direction probed by the cross-correlation
should be a relative
suppression at all separations.  The corresponding
scale factor can be evaluated at $\Delta v = 0$, where the
amplitude of $\xi^\sigma_\perp$ is known by virtue of
equation~\ref{eq_autosigma2} and the fact that $\xi_\perp\left(0\right) =
\xi_\parallel\left(0\right)$ by definition.  The 
approximate solution

\begin{equation}\label{eq_crosssigma2} 
  \xi^\sigma_\perp\left(\Delta v\right)
  \,\, \approx \,\,
  \xi_\perp\left(\Delta v\right)\,\left(1+\beta \,
  \xi_\perp\left(\Delta v\right)\right)^{\textrm{-}1},
\end{equation}

\noindent where

\begin{equation}\label{eq_beta}
  \beta 
  \,\, \equiv \,\, 
  \frac{1}{\xi^\sigma_\parallel\left(0\right)} -
  \frac{1}{\xi_\perp\left(0\right)},
\end{equation}

\noindent agrees remarkably well with results obtained using
equation~\ref{eq_crosssigma1}.  This is demonstrated in
Figure~\ref{fig_res_cross} for a representative range of redshifts and
spectral resolutions.
The slight disagreement between the two methods scales with the degree
of correlation suppression; however, the difference is
$\scriptstyle\lesssim$ 2\% for $2<z<3$, FWHM $\leq 2.5$ \AA, and $\theta>90\arcsec$.

\section{Potential Systematics}\label{sec_systematics}

Simulation of the \lya\ flux correlation is subject to a number of sources of
systematic error.  Some are either addressed by 
previous studies or may be controlled for in a limited
fashion by judicious comparison of results from the simulations listed
in Table~\ref{tab_sims}.  Others, we can only identify and
acknowledge, but not measure or correct for.  However,
the primary interest of this study is in alterations of the
correlation function due to redshift-space distortions, for which much
of this systematic uncertainty is mitigated.
It is also worth noting that while errors in the autocorrelation at small scales are
propogated to larger scales when spectral smoothing is considered,
uncertainty in the cross-correlation is only relevant at the scales
corresponding to observed QSO pair separations
($1.5\arcmin\scriptstyle\lesssim\textstyle\theta\scriptstyle\lesssim\textstyle4\arcmin$ 
in the case of Paper I).  

The effects of box length and mass resolution have already been
discussed in \S~\ref{sec_hybrid}.  For the limited scales
accessible with the \emph{q} simulation series,
our hybrid correlation
measurements appear to be largely unaffected by mass resolution and
reasonably well corrected for box length limitations with a constant
offset.  However, given the rapid decline of the correlation function
on scales affected by the small box length of \vzw, establishing
limits for the relative effect of deviations from a constant
suppression is speculative.

While the evolution of large scale structure at $z>2$ is
relatively insensitive to the cosmological parameters $\Omega_m$ and
$\Omega_\Lambda$, the adopted simulation value of $\sigma_8$ 
\citep[which
is, however, consistent with the three-year \emph{WMAP} value; ][]{2007spe}
likely does affect the correlation of the \lya\ forest.  Furthermore, the
simulations used in this study represent only a few realizations of
random fluctuation amplitudes in the early universe.  Therefore, we
cannot account for any related variance in the correlation
measurements.  Simulating the grid of amplitudes
necessary for this purpose with an SPH code is computationally
prohibitive at this time; however, see \citet{2003mcd585apj34} for a
discussion on
the alternative use of hydro-particle-mesh simulations.
In the following subsections, we address several remaining
potential sources of systematic error.

\subsection{Redshift Evolution}
                                                                                
\begin{figure} 
\begin{center}
\includegraphics[width=3.5in]{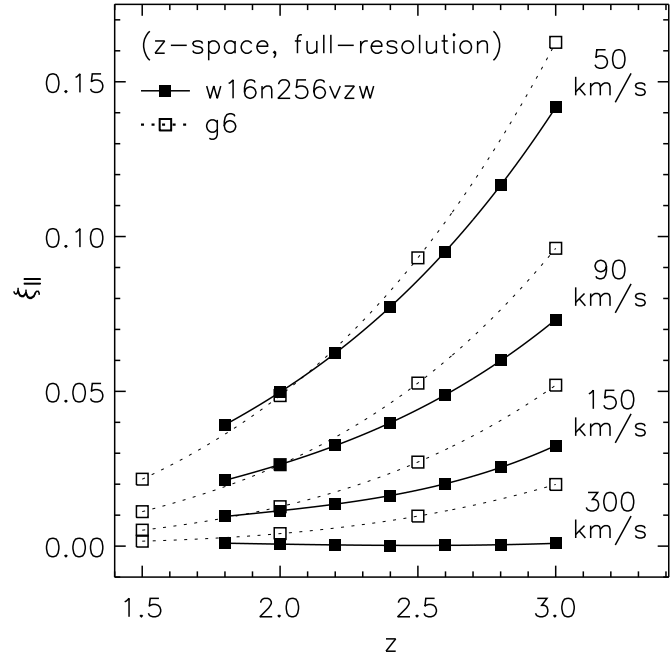}
\end{center}
\caption{The $\Delta z=0.2$ sampling of \vzw\ shows that the
  redshift-evolution of the correlation function is monotonic and smooth, allowing
  reliable interpolation with a third order polynomial (\emph{solid and
  dotted lines}).}\label{fig_zevo}
\end{figure}

The $\Delta z = 0.2$ sampling of the \vzw\ simulation was used to
verify that the redshift evolution of the autocorrelation and cross-correlation
is smooth and well-behaved over
the range of interest.  Figure~\ref{fig_zevo} illustrates
this for the case of the autocorrelation (four representative $\Delta v$
lags are shown) at full-resolution with redshift-space distortions.
A third order polynomial does an excellent job of fitting all seven
epochs, allowing for reliable interpolation at intermediate redshifts.
By extension, the same is assumed for the more coarsely sampled
($\Delta z = 0.5$) \g6 simulation.

\subsection{Metals}

The simulated spectra generated for this study include absorption
from \ion{H}{1} only; however, the \lya\ forest in observed spectra is
contaminated by metal lines.  
Associated metals can introduce features into the correlation function
at the velocity difference,

\begin{equation}
  \Delta v \simeq c \, \Delta \lambda \, / \, \lambda,
\end{equation}

\noindent between their absorption and that of \lya\ from the same gas.
Indeed, \citet{2006mcd163apjs80} found enhanced correlation at $\Delta
v \simeq 2270$ \kms\ due to \ion{Si}{3} at rest wavelength
1206.50 \AA; however, no other metal correlations were detected.
More to the point, no metals in the IGM have known wavelengths closer
to that of \lya\ than \ion{Si}{3}; thus increased correlation from
associated metals is not a concern for the velocity
scales relevant to this study. 
Similarly, the velocity splitting of the \ion{Si}{4} doublet ($\Delta
v \simeq 1930$ \kms) lies beyond our range of consideration, while 
\citet{2006mcd163apjs80}
found no evidence of a correlation feature at $\Delta v \simeq 500$ \kms\
corresponding to the \ion{C}{4} doublet.

A third potential source of increased correlation
is the clustering of metals themselves.  This effect cannot be
accounted for in the simulated spectra for two reasons.  First,
unassociated metals sparsely populating the \lya\ forest arise from
gas at lower redshifts, beyond the epoch for which the simulations
were run in some cases.  Second, although the {\sc vzw} wind model
(see \S~\ref{sec_wind})
has been shown to reproduce the overall mass density and absorption
line properties of \ion{C}{4} well, the simulations do not yet
accurately reflect the clustering properties of metals.  Fortunately, 
clustering of metals is not expected to significantly affect the \lya\
flux correlation function; absorption from even the most abundant metals is
2--4 orders of magnitude less than \ion{H}{1}
\citep{2003sch596apj768,2003fry281assl231}.

\subsection{Mean Flux Decrement Uncertainty}\label{sec_da}

The mean flux decrement \citep{1982oke255apj11O},

\begin{equation}
  D_{\!A}\,=\,1- \langle f \rangle,
\end{equation}

\noindent where $\langle f \rangle$ is the mean of the transmitted flux 
(observed flux divided by the unabsorbed continuum flux)
in the \lya\ forest can be reliably tuned to high precision in
simulated spectra (recall \S~\ref{sec_spectra}); however, this is only as accurate as the 
observationally
determined value.  Measurement of \da\ from real spectra is
complicated by the difficult step of estimating the continuum of
emitted flux from the background light source
(conveniently defined as unity in simulated spectra).
For low-resolution spectra, the
continuum has generally been extrapolated from redward of the \lya\
forest, assuming a power-law.  This technique will not likely be
accurate for an individual spectrum; however, the significant
uncertainties are assumed to be mitigated for a sufficiently large sample.
In the case of higher resolution spectra, a smooth continuum is fit to
regions free of obvious absorption.  While individually tailored,
residual absorption will almost certainly result in artificially low
continuum placement (corresponding to underestimated absorption)
unless this bias can be adequately modelled.

\begin{figure} 
\begin{center}
\includegraphics[width=3.5in]{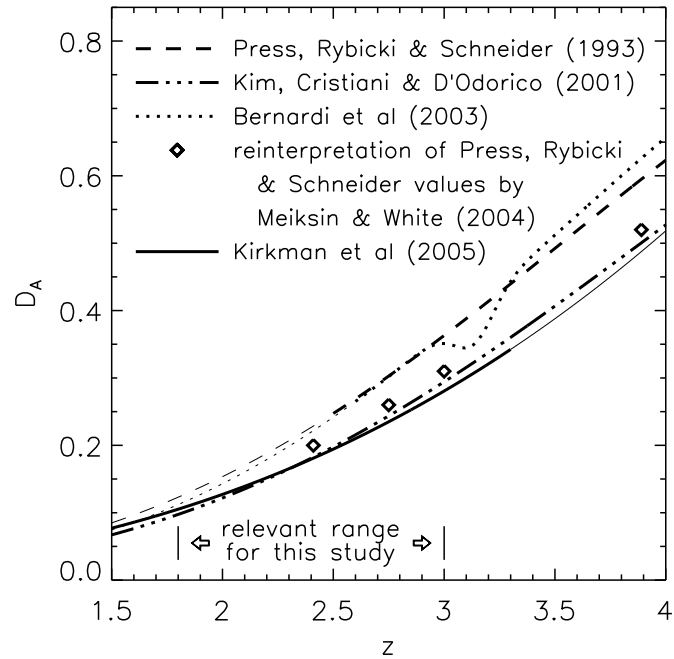}
\end{center}
\caption{Mean flux decrement, \da, of the \lya\ forest remains
  observationally uncertain.}\label{fig_da}
\end{figure}

Numerous measurements of \da\ have been made during the past two
decades (see \citet{1998rau36araa267}, \citet{2004mei350mnras1107}, and
references therein).  The few that also determined its evolution as a
function of redshift are compared in Figure~\ref{fig_da}, where the
thick lines represent the redshift range of the data used.
Consistent with the above discussion, \citet{1993pre414apj64} extrapolated the continuum for 29
quasars and obtained

\begin{equation}
  D_{\!A}^{P93}(z) = 1 - e^{-0.0037\,(1+z)^{3.46}}, \label{eq_p93}
\end{equation}

\noindent whereas continuum fits to echelle-resolution spectra by
\citet{2001kim373aap757} and \citet{2005kir360mnras1373} yielded
significantly lower values (less absorption).  The latter found 

\begin{equation}
  D_{\!A}^{K05}(z) = 0.0062\,(1+z)^{2.75} \label{eq_k05}
\end{equation}

\noindent and claimed errors of less than 1\% based on
tests using artificial 
spectra.  A reevaluation of the \citet{1993pre414apj64} results by 
\citet{2004mei350mnras1107} reported much better agreement with the
high resolution studies; however, \citet{2003ber125aj32} similarly
extrapolated the continua for a sample of 1061 quasars and produced
results very similar to the original \citet{1993pre414apj64} values.
The mean flux decrement remains observationally uncertain. 

\begin{figure} 
\begin{center}
\includegraphics[width=3.5in]{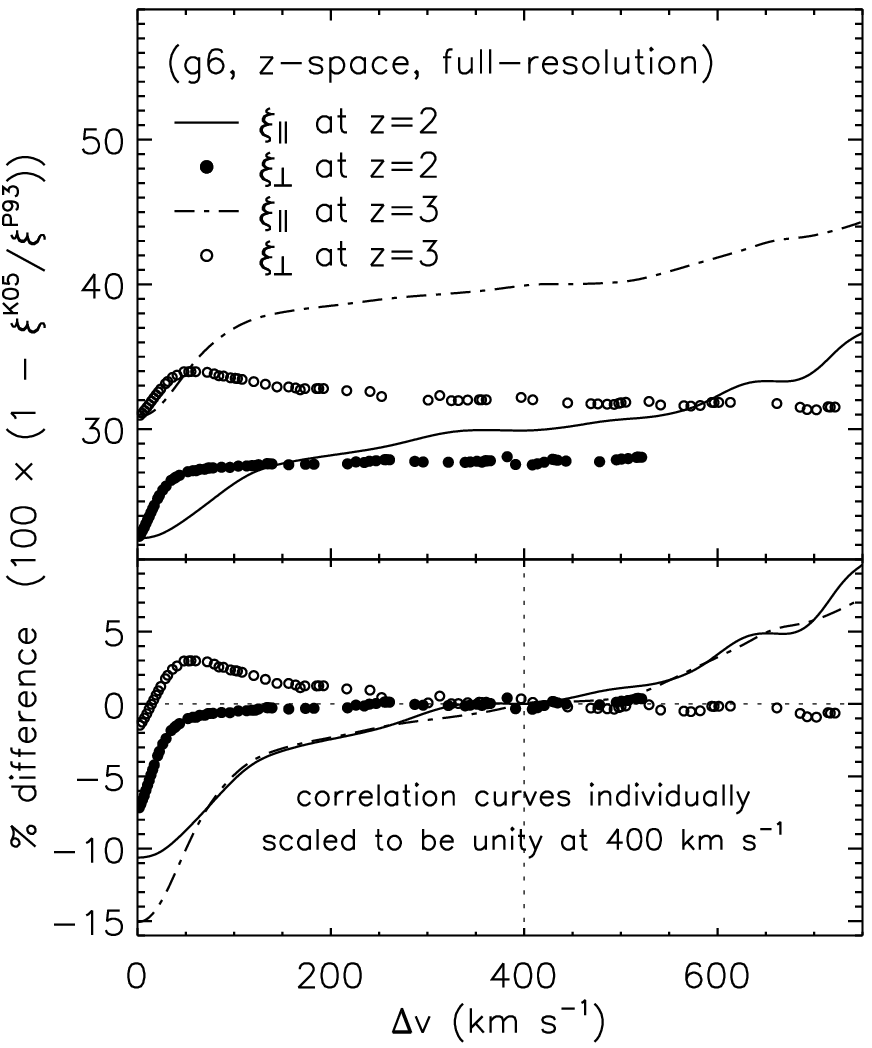}
\end{center}
\caption{Amplitude and shape of the 
  \lya\ flux correlation function are sensitive 
  to the mean flux decrement, as evidenced by the relative
  difference in $\xi$ for simulated spectra tuned to match the values
  from \citet[][P93]{1993pre414apj64} and 
  \citet[][K05]{2005kir360mnras1373}.}\label{fig_meanflux}
\end{figure}

Unfortunately, as has been shown previously for matter power
spectrum measurements made with the \lya\ forest 
\citep{2002cro581apj20,2003zal590apj1,2003sel342mnras79},
both the amplitude and shape of the correlation function are sensitive 
to \da.  Figure~\ref{fig_meanflux}
shows the percent difference in the correlation function for simulated
spectra tuned to have the mean flux decrement prescribed by either
equation~\ref{eq_p93} or equation~\ref{eq_k05}.
In the bottom panel of Figure~\ref{fig_meanflux}, the correlation functions compared have been 
arbitrarily scaled to
unity at 400 \kms\ in order to mitigate differences solely in
amplitude.  We have addressed this systematic uncertainty by carrying
out our analysis using the mean flux decrement values of both
\citet{1993pre414apj64} and \citet{2005kir360mnras1373}.  Unless
otherwise stated, results from the former are used in the figures
throughout this paper (where this choice is secondary to other effects
being considered).

\subsection{Spectral Resolution}

\begin{figure} 
\begin{center}
\includegraphics[width=3.5in]{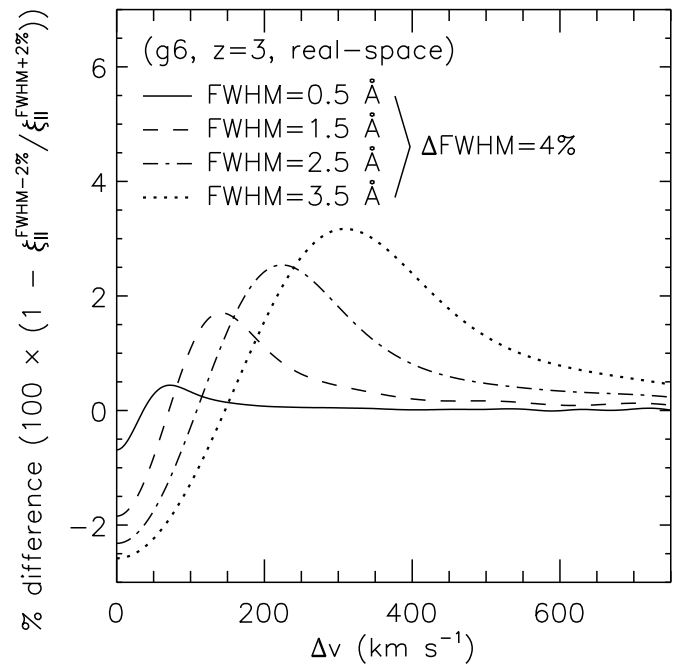}
\end{center}
\caption{Relative difference in autocorrelation
  corresponding to a 4\% change in spectral resolution FWHM.}\label{fig_res_err}
\end{figure}

Section~\ref{sec_resolution} discussed how to account for arbitrary spectral
resolution when using the correlation functions computed at
full-resolution.  Here we consider in greater detail the sensitivity
of the correlation function to small changes in spectral resolution
(or uncertainty in that parameter).
Figure~\ref{fig_res_err} shows the relative difference
in autocorrelation corresponding to a 4\% change ($\Delta $FWHM $=0.1$\AA\
for FWHM $=2.5$\AA) in resolution.  This difference scales roughly
linearly for larger $\Delta $FWHM and is less for the cross-correlation.
Thus, treating data with FWHM $=2.3$ and $2.5$ \AA\ as having the same
resolution introduces an error of up to $5\%-6\%$.

\subsection{Wind Model}\label{sec_wind}

Prescriptions for galactic winds, which transport processed gas from within
galaxies to the surrounding IGM, are relatively
new additions to cosmological simulations.  \citet{2003spr339mnras312}
incorporated a constant wind ({\sc cw}) model 
in order to reduce the amount of gas available for star
formation in galaxies.  Essentially, a fraction of the gas particles,
dictated by the current star formation rate and a relative mass loading factor $\eta$,
are ejected from a galaxy via superwinds.  They then travel without
hydrodynamic interaction at a constant velocity $v_{wind}$ until the
SPH density falls below 10\% of the critical density for multi-phase
collapse.  Based on earlier simulation work by \citet{2001agu561apj521}
and observations from \citet{1999mar513apj156} and \citet{2000hec129apjs493}, the
two free model parameters were set at $v_{wind} = 484$ \kms\
and $\eta = 2$.  This yields broad agreement with observations of
the stellar mass density at z=0; however, the wind velocity is
unphysically large for small galaxies, and \citet{2006opp373mnras1265}
found that \ion{C}{4} is overproduced in the IGM compared to observed
$\Omega_{\rm{C\,IV}}$ data.  These authors also note that the
{\sc cw} model does not converge well with resolution.  That is, in
higher-resolution simulations which resolve small galaxies earlier,
the winds turn on earlier and heat the IGM in excess of
lower-resolution simulations. 

\citet{2006opp373mnras1265} also investigated several, more sophisticated prescriptions
 for galactic outflow, contrasting their effect on the IGM and
 comparing the results to observational data. The most successful models were variants of 
momentum-driven winds.  In the case of {\sc vzw}, the wind speed, $v_{wind} = 3\, \sigma
\sqrt{f_L-1}$, and the mass loading factor, $\eta =
\sigma_o\,\sigma^{\textrm{-}1}$, both scale as the galaxy velocity dispersion,
$\sigma=\sqrt{-\Phi/2}$.  Here, $\Phi$ is the gravitational potential, 
and $f_L=f_{L,\sun}\times10^{\,0.0029 \,(log Z+9)^{2.5}\,+\,0.417694}$ is the galaxy
luminosity in units of its critical luminosity.  The free parameter
 $\sigma_o$ was chosen to be 300 \kms, corresponding to a
 Salpeter initial mass function and a typical starburst spectral energy distribution, and $f_{L,\sun}$ was
 allowed to vary randomly in the range $1.05-2$ as observed by
 \citet{2005rup160apjs115}.  Unlike the {\sc cw} wind model, {\sc vzw} was
 shown to non-trivially reproduce a wide range of \ion{C}{4}
 absorption observations.

\begin{figure} 
\begin{center}
\includegraphics[width=3.5in]{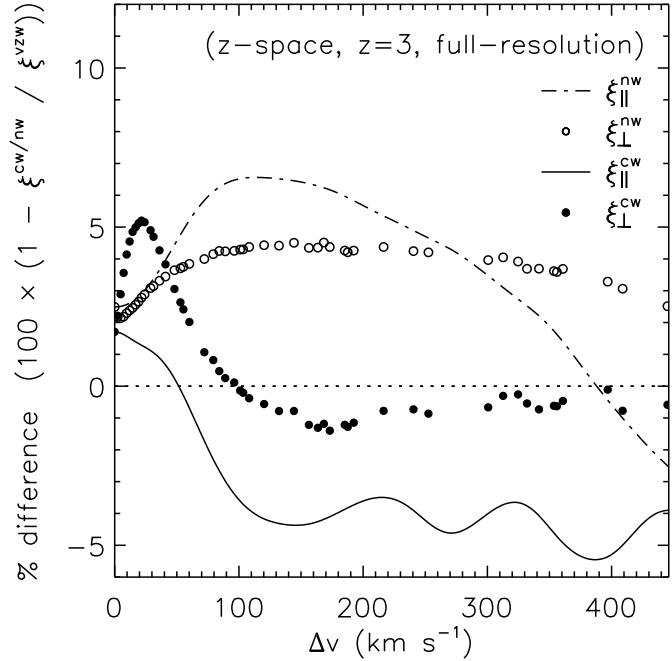}
\end{center}
\caption{Percent difference in correlation between \vzw\ and
  \emph{w16n256}[\emph{cw}/\emph{nw}] indicates that \lya\
  correlation measurements are marginally affected by the
  prescription for galactic outflow. The preferred momentum-driven
  wind model {\sc vzw} differs from the
  older constant wind model {\sc cw} at $z=3$ and on scales larger than $\sim\,$50
  \kms\ by $\scriptstyle\lesssim$ 1\% for the cross-correlation and
  $\scriptstyle\lesssim$ 4\%
  for the autocorrelation.  This increases to $<\,5$\% and $<\,7$\%,
  respectively, when compared to no winds at all ({\sc nw}).}\label{fig_wind}
\end{figure}

While more detailed studies of the effects of galactic winds on the
\lya\ forest have been carried out \citep{2002cro580apj634,
  2004des350mnras879, 2005mcd360mnras1471}, our primary interest is in
investigating how the different wind models included in the \g6\
({\sc cw}) and \vzw\ ({\sc vzw}) simulations might affect our \lya\ forest
flux correlation measurements.
The \cw\ and \nw\
simulations are identical to \vzw\, with the exception of their
wind models.  As their nomenclature indicates, the former incorporates
the {\sc cw} model, while the latter includes no winds at all ({\sc nw}).
Slight differences in the flux distribution caused by the inclusion of
winds were mitigated by the rescaling of opacities described in
\S~\ref{sec_spectra}.  Although the three simulations are identically
affected by box length limitations, corrections were applied as
described in \S~\ref{sec_hybrid} so that meaningful comparisons could be
made of correlation curves that otherwise cross zero in the region of interest.
Figure~\ref{fig_wind} shows the percent difference between the
correlation values obtained from each of these two simulations
and those from \vzw\ (in redshift-space, at $z=3$, and at
full-resolution). 
The \cw\ and \vzw\ results differ by $\scriptstyle\lesssim$ 1\%
and $\scriptstyle\lesssim$ 4\%, for the cross-correlation and autocorrelation
respectively, on
scales larger than $\sim\,50$ \kms, indicating that our correlation
measurements for \g6\ and \vzw\ are only marginally affected by the
use of different wind models.  Furthermore, while we assume that
inclusion of the currently preferred wind model yields more accurate
results than neglecting galactic winds altogether, the \nw\ and
\vzw\ comparison demonstrates that these two extremes represent a
difference of only $\scriptstyle\lesssim$ 7\%.

\section{Implications For The AP Test}\label{sec_ap}

\subsection{Signal-To-Noise Ratio}

Cross-correlation measurements were repeated
with varying degrees of Gaussian noise added to the individual simulated
(\vzw) spectra.  Although this has no effect on the mean
correlation values (which have been averaged over many lines of
sight), signal-to-noise ratio ($S/N$) does affect the dispersion of those values.
However, even for relatively low $S/N$, the
corresponding increase in $\sigma_{\xi}$ is negligible relative to the
intrinsic variation in $\xi$ between different lines of sight.  The
latter scales inversely with path length, but even for the entire \lya\
forest redward of \lyb\ absorption, the difference in
$\sigma_{\xi}$ between $S/N=5$ and $S/N=\infty$ is 
$\scriptstyle\lesssim$ 2\% for
$2<z<3$, resolution FWHM $\le2.5$\AA, and $0<\theta<300$ arcseconds.
This is in agreement with the assertion by \citet{2003mcd585apj34} that
only moderate quality data is needed for a large number of quasar
pairs to carry out the Alcock-Paczyn\'{n}ski test.
The $S/N$ requirements of observed spectra are dictated not
by correlation measurements, but by the needs of reliable continuum fitting.

\subsection{Continuum Errors and \da\ Variance}

Errors in fitting the continua of observed QSO spectra can affect
calculation of the 
mean flux decrement (recall \S~\ref{sec_da}) as well as correlation measurements.
Comparison of \da\ for a particular spectrum to the expected
mean flux decrement might be used, in principle, to constrain
systematic errors in the determination of the continuum.  
However, genuine variation in \da\ arises naturally
between lines of sight (decreasing with increasing path length) due to
finite sampling of the local large scale structure.
Simulated spectra provide an opportunity to quantify the expected
distribution of mean flux decrement measurements
in the absence of continuum fitting errors.

Table~\ref{tab_davar} provides the variance in \da\ as a function of
redshift and path length (in units of $h^{\textrm{-}1}$ comoving Mpc) for 
\vzw\ and \g6.  Although some validation is given by the
general agreement between the two simulations, the \g6\ results are
systematically lower than those for \vzw\ (the percent difference
increases from approximately 1\% to 9\% at $z=3$ and 2,
respectively).  If the difference was dominated by the
diversity in large scale structure contained within the different
simulation volumes, one would expect the \g6\ variances to be larger.
Since this is not the case, we presume that the differences primarily
reflect the greater mass resolution of the \vzw\ simulation (note
that this is consistent with the difference increasing monotonically
as the fraction of pixels in low density regions increases at lower
redshift).  At $z=2.2$, the standard deviation in \da\ for a
path length of $\Delta z=0.2$ is $\sigma_{D_{\!A}}\!\simeq0.017$,
corresponding to 9.0\% and 12.1\% of the 
\da\ value from \citet{1993pre414apj64} and 
\citet{2005kir360mnras1373}, respectively.  This decreases
to $\sigma_{D_{\!A}}\simeq0.010$ (5.5\% and
7.3\%) for $\Delta z=0.545$, the path length of the full ``pure''
\lya\ forest (redward of the onset of \lyb\ absorption).

\subsection{Anisotropy Corrections}

The primary goal of this work is to model anisotropies 
in the observed \lya\ forest correlation function, facilitating a new
application of the AP test using spectra of QSO pairs (such as those presented in
Paper I).
To this end, we have computed the autocorrelation and cross-correlation in both
real and redshift-space, investigated potential sources of systematic
error, and considered the impact of spectral smoothing.
Our full-resolution, hybrid correlation measurements are provided in
Tables~\ref{tab_cross_p93}-\ref{tab_auto_k05} (complete versions of
the stubs included here can be found in the electronic edition of \apj\ or upon request) for the mean flux decrements of both
\citet{1993pre414apj64} and \citet{2005kir360mnras1373}.  Note that
the velocity scales are redshift dependent, so a unitless
parameterization (first column) is used which is not the same for the
autocorrelation and cross-correlation.

Implementation of the AP test itself is nontrivial and the subject of
Paper III in this 
series.  However, we conclude by outlining a scheme for the use of
these simulation results that mitigates the systematic uncertainty discussed in
\S~\ref{sec_systematics}.  To reiterate, the correlation function of
the resolved \lya\ forest is isotropic in real-space, and adjusting the angular
diameter distance until cross-correlation measurements (the data) agree
with the autocorrelation (the model) yields the correct cosmology.

\begin{figure} 
\begin{center}
\includegraphics[width=7.25in]{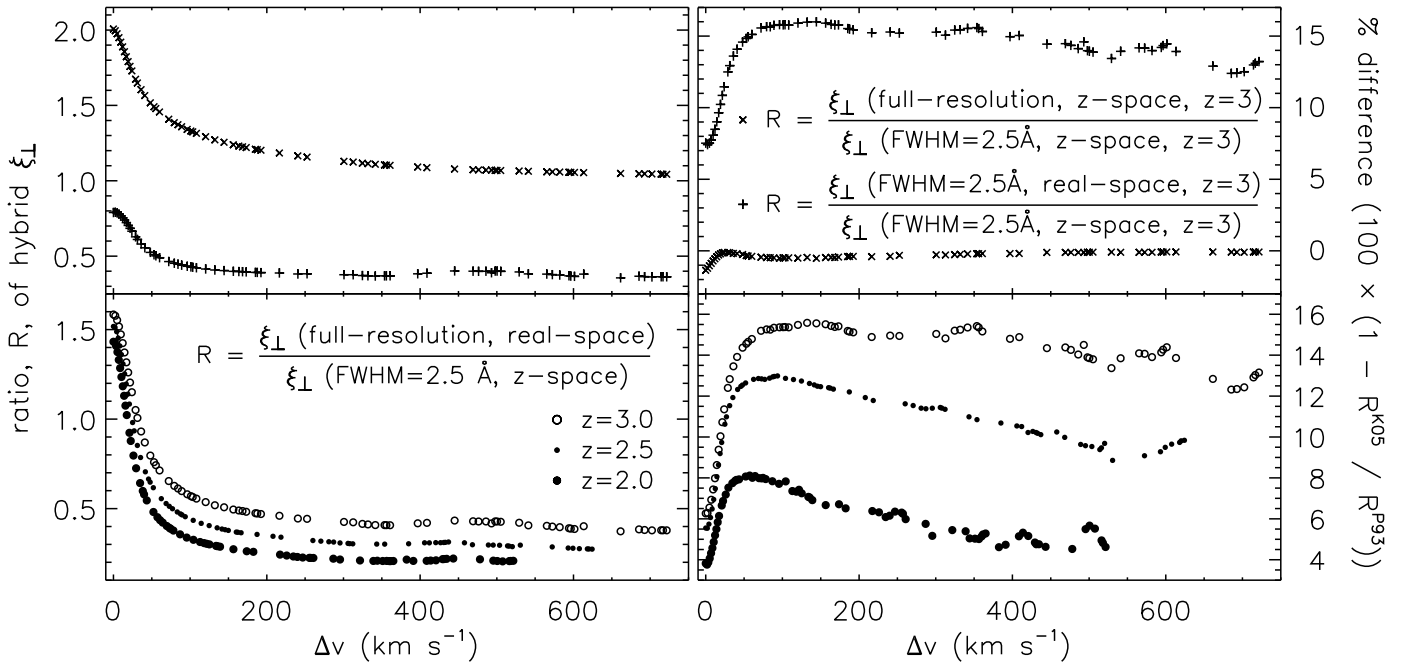}
\end{center}
\caption{Ratios ($R$) in the left panels show the
  individual effects on the cross-correlation of spectral 
  smoothing and redshift-space distortions at $z=3$ (\emph{top}) and their
  combined impact at $z=2$, 2.5, and 3 (\emph{bottom}).  The relative difference
  between using the mean flux decrements measured by
  \citet[][P93]{1993pre414apj64} or
  \citet[][K05]{2005kir360mnras1373} are given in the right panels.}
\label{fig_crossratio}
\end{figure}

Figure~\ref{fig_crossratio} shows the effects 
of redshift-space distortions and spectral
smoothing on the cross-correlation 
(\emph{top left}).  These can be accounted for in observed
cross-correlation measurements by applying the ratio
of the full-resolution, real-space simulated
cross-correlation divided by its counterpart for smoothed data (using
equations~\ref{eq_crosssigma2} and \ref{eq_beta})
in redshift-space (bottom left panel of Figure~\ref{fig_crossratio}).
This correction requires adopting an angular diameter distance
and, therefore, must be applied independently for each cosmology
considered.  Using the ratio of simulation results allows for partial
cancellation of systematic errors.  The right panels of
Figure~\ref{fig_crossratio} show the relative difference in these
corrections between using the mean flux decrement of
\citet{1993pre414apj64} or \citet{2005kir360mnras1373}.  While
still a significant source of systematic uncertainty, the impact of
\da\ on the correlation ratio is reduced relative to the correlation
function itself (recall Figure~\ref{fig_meanflux}).

Until sufficient high-resolution (echelle) data exists for reliable
determination of the autocorrelation (many lines of sight are needed
to compensate for significant variance), simulated measurements provide
the only reasonably continuous model.  However, more abundant observational data
obtained at lower resolution can be used to correct systematic error
in the simulation data.  This is accomplished by smoothing the
full-resolution, redshift-space correlation curve (recall that the
$z=3$ hybrid autocorrelation is shown in Figures~\ref{fig_zspace} and
\ref{fig_res}) as appropriate
(using equations~\ref{eq_smoothing} and \ref{eq_autosigma2}) and fitting
it to the observed data. The same corrections can then be applied to the
simulated full-resolution, real-space autocorrelation model, which is
not affected by the discussed anisotropies.

\section{Summary}\label{sec_summary}

Using cosmological hydrodynamic simulations, we have modelled
the \lya\ flux autocorrelation and zero-lag cross-correlation in both
real-space and redshift space at $1.8<z<3$.
 Mock \lya\ flux absorption spectra were generated from eight SPH
   simulations with and without inclusion of redshift-space
   distortions caused by Hubble expansion, bulk flows, and thermal
   broadening. The simulations considered (\emph{w16n256vzw}
   at $1.8 \le z \le 3.0$, \g6\ at $1.5 \le z \le 3.0$, and
   \emph{w16n256cw}, \emph{w16n256nw}, \qi, \qii, \qiii, and \qiv\ at
   $z=3$) primarily differ in their size, mass resolution,
   and prescription for galactic outflow. The lines of sight
   through each simulation box were selected such that different
   pairings form 73 unique transverse separations spanning the range
   $0-5$ arcminutes.  Our analysis is summarized below.

1) Autocorrelation and zero-lag cross-correlation measurements were
   computed from the extracted spectra for both real-space and
   redshift-space and for the mean flux decrement values reported by
   both \citet{1993pre414apj64} and \citet{2005kir360mnras1373}.  The 
   difference in the autocorrelation and cross-correlation
   corresponding to this observationally uncertain parameter was found
   to be $20\%-45\%$ and $20\%-35\%$, respectively, affecting both the
   shape and amplitude.

2) Convergence of the simulated \lya\ flux correlation as a function of mass
   resolution was tested at $z=3$ for the {\sc Gadget} code using the
   \emph{q}-series simulations (which identically evolve different numbers of
   particles within boxes of equal volume).  The difference in autocorrelation and
   cross-correlation between \qiii\
   ($m_{gas}=3.72\times10^6~h^{\textrm{-}1} ~M_{\sun}$)
   and \qiv\ ($m_{gas}=1.10\times10^6~h^{\textrm{-}1} ~M_{\sun}$) is less
   than 3\% on all scales.  

3) The \emph{q}-series was also used to characterize
   the effect of insufficient mass resolution in \g6\ and, indirectly,
   the effect of the inadequate simulation volume of \emph{w16n256vzw}.  In 
   order to correct for these limitations of current simulations,
   hybrid correlation curves were then formed by splicing together
   those from \emph{w16n256vzw} and \g6\ at $\Delta v = 445$ \kms.  At smaller
   velocities, the hybrid correlation is equal to that of
   \emph{w16n256vzw} plus a constant boxsize correction (in the case
   of the cross-correlation, this is slightly modified to preserve the
   boundary condition at $\Delta v=0$).  At larger
   velocities, where the effects of mass resolution were projected to
   be insignificant, the hybrid correlation is provided by that of
   \g6\ without alteration.

4) An approximate solution is presented for obtaining the
   zero-lag cross-correlation corresponding to arbitrary spectral resolution
   directly from the zero-lag cross-correlation computed at
   full-resolution (an exact solution is available in the case of the
   autocorrelation).  This approximation is good to within 2\% for the
   relevant redshift range at
   velocity differences corresponding to angular separations greater
   than 90 arcseconds. 

5) The effects of three prescriptions for galactic outflow on the
   \lya\ flux correlation were investigated with the 
   \emph{w16n256vzw},
   \emph{w16n256cw}, and \emph{w16n256nw}
   simulations.  The difference between the
   preferred variable-momentum wind model ({\sc vzw}, used for \emph{w16n256vzw}) and the
   older constant wind model ({\sc cw}; used for \g6) was found to be 
   $\scriptstyle\lesssim$ 1\% and $\scriptstyle\lesssim$ 4\% at 
   scales larger than $\approx 50$ \kms\ for the cross-correlation and autocorrelation
   respectively.  The corresponding difference between {\sc vzw} and no winds at all
   increases to only $< 5\%$ and $< 7\%$.

6) For an adopted mean flux decrement, the variance from one line of sight to
   another was computed as a function of redshift and path length.
   At $z=2.2$, the standard deviation in \da\ for a
   path length of $\Delta z=0.2$ is $\sigma_{D_{\!A}}\!\simeq0.017$,
   corresponding to 9.0\% and 12.1\% of the 
   \da\ value from \citet{1993pre414apj64} and 
   \citet{2005kir360mnras1373}, respectively.

7) Aside from those sources of systematic error already summarized
   above, we find that redshift evolution of the \lya\ flux correlation 
   is sufficiently sampled for reliable interpolation 
   and argue that absorption from metals is insignificant.  
   The evolution of large scale structure at $z>2$ is not sensitive to the
   values for the cosmological parameters $\Omega_m=0.3$ or
   $\Omega_\Lambda=0.7$ assumed by the simulations considered here, 
   and $\sigma_8=0.9$ is consistent with the three-year \emph{WMAP} value.  
   Systematic error associated with
   variance of random fluctuation amplitudes in the early universe or
   deviations from a constant offset due to finite boxsize cannot be
   addressed with currently available simulations.

8) Correcting for anisotropies due to
   redshift-space distortions and spectral smoothing
   with ratios of the correlation
   measurements allows for significant reduction in systematic error.
   The maximum difference between using the mean flux
   decrements of either \citet{1993pre414apj64} or
   \citet{2005kir360mnras1373}
   (the dominant source of uncertainty) decreases to
   $8\%-16\%$ at $2<z<3$, and presumably the true value is intermediate.  
   We describe a simple scheme for implementing our results, while mitigating
   systematic errors, in the context of a future
   application of the AP test using observations of the \lya\ forest
   in pairs of QSOs.  

\acknowledgements

This study would not have been possible without significant access to
the Beowulf computer cluster (Mendeleyev) at Steward Observatory
and the corresponding generosity of Dave Arnett,
Adam Burrows, Daniel Eisenstein, Phil Pinto, and Dennis Zaritsky.  Additionally, we
owe gratitude to Jeff Fookson and Neal Lauver for administering the
cluster and supporting this work.  We thank Patrick McDonald, Daniel Eisenstein, Martin
Pessah, Chi-Kwan Chan, Volker Springel, Lars
Hernquist, and Lei Bai for helpful conversations along the way.

\clearpage 

\clearpage


\begin{deluxetable}{l l r@{}l l c c c c c r}
\tablecolumns{11}
\tablewidth{0pc}
\tablecaption{Simulation Properties\label{tab_sims}}
\tablehead{
\multicolumn{2}{c}{Simulation} & \multicolumn{2}{c}{$L$\tablenotemark{a}} &
\colhead{$N_p$\tablenotemark{b}} & \colhead{$m_{gas}$\tablenotemark{b}} & 
\colhead{$\epsilon\,$\tablenotemark{a}} &
\colhead{Wind$\,$\tablenotemark{c}} & \multicolumn{2}{c}{Redshift} & \colhead{$N_s$}\\ 
\cline{1-2}
\cline{9-10}
\colhead{Name} & \colhead{Alias} & \multicolumn{2}{c}{($h^{-1}$ Mpc)} &
\colhead{} & \colhead{($h^{-1} $M$_{\sun}$)} & 
\colhead{($h^{-1}$ kpc)} & \colhead{} & \colhead{Range} &
\colhead{$\Delta z$} & \colhead{}}
\startdata
\qi                & \nodata &  10&    & $2\times64^3$  & $4.24\times10^7$ & 6.25 & {\sc cw}  & 3.0       & \nodata & 30000 \\ 
\qii               & \nodata &  10&    & $2\times96^3$  & $1.25\times10^7$ & 4.17 & {\sc cw}  & 3.0       & \nodata & 30000 \\ 
\qiii              & \nodata &  10&    & $2\times144^3$ & $3.72\times10^6$ & 2.78 & {\sc cw}  & 3.0       & \nodata & 30000 \\ 
\qiv               & \nodata &  10&    & $2\times216^3$ & $1.10\times10^6$ & 1.85 & {\sc cw}  & 3.0       & \nodata & 30000 \\ 
\emph{w16n256nw}   & \nw     &  16&    & $2\times256^3$ & $2.71\times10^6$ & 1.25 & {\sc nw}  & 3.0       & \nodata & 20000 \\
\emph{w16n256cw}   & \cw     &  16&    & $2\times256^3$ & $2.71\times10^6$ & 1.25 & {\sc cw}  & 3.0       & \nodata & 20000 \\
\emph{w16n256vzw}  & \vzw    &  16&    & $2\times256^3$ & $2.71\times10^6$ & 1.25 & {\sc vzw} & $1.8-3.0$ & 0.2     & 20000 \\
\g6                & \nodata & 100&    & $2\times484^3$ & $9.79\times10^7$ & 5.33 & {\sc cw}  & $1.5-3.0$ & 0.5     &  3000 \\
\enddata

\tablecomments{The columns (left to right) are: simulation name and abbreviation, box
  length, total number of particles, gas particle mass, equivalent
  Plummer gravitational softening length, wind model, redshift range
  and interval, and number of sets of lines of sight extracted along
  each of the three principal axes.}

\tablenotetext{a}{These physical scales are given in comoving coordinates.}
\tablenotetext{b}{The total number of particles is evenly divided
  between dark matter and gas.  Therefore, the dark matter
  particle mass, $m_{dm}$, is simply the gas particle mass
  scaled by $(\Omega_m-\Omega_b) / \Omega_b$.}
\tablenotetext{c}{The simulations have either no prescription for
  galactic outflow ({\sc nw}), a constant wind ({\sc cw}) model, or momentum-driven
  winds ({\sc vzw}).  See \S~\ref{sec_wind} for details.}

\end{deluxetable}

\begin{deluxetable}{c c c}
\tablecolumns{3}
\tablewidth{0pc}
\tablecaption{\da\ Variance\label{tab_davar}}
\tablehead{
\colhead{$z$} &
\multicolumn{2}{c}{$\sigma^2_{D_{\!A}}\times l\,$\tablenotemark{a}}\\
\cline{2-3}
\colhead{} & \colhead{\vzw} & \colhead{\g6}}
\startdata
1.5 & \nodata  & 0.024144 \\ 
1.8 & 0.041109 & \nodata  \\ 
2.0 & 0.048716 & 0.044152 \\ 
2.2 & 0.056126 & \nodata  \\ 
2.4 & 0.063490 & \nodata  \\ 
2.5 & \nodata  & 0.064115 \\
2.6 & 0.069438 & \nodata  \\
2.8 & 0.074085 & \nodata  \\
3.0 & 0.076494 & 0.075647 \\ 
\enddata

\tablenotetext{a}{The variable $l$ is the path length in $h^{-1}$ comoving Mpc}

\end{deluxetable}


\begin{deluxetable}{crrrrrrrrrrrrrrr}
\tablecolumns{16}
\tabletypesize{\tiny}
\tablewidth{7.25in}
\tablecaption{Cross-Correlation (\citet{1993pre414apj64} Mean Flux Decrement)\label{tab_cross_p93}}
\tablehead{
\colhead{$\Delta v$\tablenotemark{a}} & \multicolumn{7}{c}{$10\times\xi_\perp\!\left(\Delta v\right)$ in real-space} & 
\colhead{} & \multicolumn{7}{c}{$10\times\xi_\perp\!\left(\Delta v\right)$ in \emph{z}-space}\\
\cline{2-8}
\cline{10-16}
\colhead{($\delta v_\perp\!\left(z\right)$)} & 
\colhead{$z=1.8$} & \colhead{2.0} & \colhead{2.2} & \colhead{2.4} & 
\colhead{2.6} & \colhead{2.8} & \colhead{3.0} & 
\colhead{} & 
\colhead{$z=1.8$} & \colhead{2.0} & \colhead{2.2} & \colhead{2.4} & 
\colhead{2.6} & \colhead{2.8} & \colhead{3.0}}
\startdata
0000 &  0.4315 &  0.5914 &  0.7994 &  1.0641 &  1.4036 &  1.8199 &  2.3333 & &  0.7489 &  0.9812 &  1.2594 &  1.5876 &  1.9748 &  2.4281 &  2.9558\\
0001 &  0.4263 &  0.5853 &  0.7917 &  1.0541 &  1.3907 &  1.8026 &  2.3094 & &  0.7302 &  0.9614 &  1.2392 &  1.5660 &  1.9499 &  2.3995 &  2.9216\\
0002 &  0.4151 &  0.5702 &  0.7718 &  1.0275 &  1.3555 &  1.7566 &  2.2504 & &  0.7089 &  0.9364 &  1.2103 &  1.5316 &  1.9089 &  2.3508 &  2.8630\\
0003 &  0.3997 &  0.5493 &  0.7436 &  0.9903 &  1.3067 &  1.6933 &  2.1686 & &  0.6870 &  0.9090 &  1.1772 &  1.4915 &  1.8593 &  2.2893 &  2.7876\\
$\downarrow$ & & & & & & & & & & & & & & & \\
0300 &  0.0078 &  0.0107 &  0.0176 &  0.0230 &  0.0238 &  0.0237 &  0.0461 & &  0.0489 &  0.0555 &  0.0654 &  0.0840 &  0.1030 &  0.1158 &  0.1270\\
\enddata
\tablenotetext{a}{74 $\Delta v$ values given in units of $\delta v_\perp\!\left(z\right) \equiv 0.67455\,z + 0.38896$ km s$^{-1}$ (or, alternatively, arcseconds)}
\tablecomments{These are the hybrid correlation 
values for full-resolution. [The complete version of this table can be found in the electronic edition of \apj\ or upon request.]}
\end{deluxetable}


\begin{deluxetable}{crrrrrrrrrrrrrrr}
\tablecolumns{16}
\tabletypesize{\tiny}
\tablewidth{7.25in}
\tablecaption{Autocorrelation (\citet{1993pre414apj64} Mean Flux Decrement)\label{tab_auto_p93}}
\tablehead{
\colhead{$\Delta v$\tablenotemark{a}} & \multicolumn{7}{c}{$10\times\xi_\parallel\!\left(\Delta v\right)$ in real-space} & 
\colhead{} & \multicolumn{7}{c}{$10\times\xi_\parallel\!\left(\Delta v\right)$ in \emph{z}-space}\\
\cline{2-8}
\cline{10-16}
\colhead{($\delta v_\parallel\!\left(z\right)$)} & 
\colhead{$z=1.8$} & \colhead{2.0} & \colhead{2.2} & \colhead{2.4} & 
\colhead{2.6} & \colhead{2.8} & \colhead{3.0} & 
\colhead{} & 
\colhead{$z=1.8$} & \colhead{2.0} & \colhead{2.2} & \colhead{2.4} & 
\colhead{2.6} & \colhead{2.8} & \colhead{3.0}}
\startdata
0000 &  0.4315 &  0.5914 &  0.7994 &  1.0641 &  1.4036 &  1.8199 &  2.3333 & &  0.7489 &  0.9812 &  1.2594 &  1.5876 &  1.9748 &  2.4281 &  2.9558\\
0001 &  0.4250 &  0.5837 &  0.7902 &  1.0530 &  1.3902 &  1.8036 &  2.3135 & &  0.7478 &  0.9797 &  1.2572 &  1.5847 &  1.9709 &  2.4231 &  2.9493\\
0002 &  0.4097 &  0.5649 &  0.7672 &  1.0249 &  1.3558 &  1.7616 &  2.2622 & &  0.7445 &  0.9750 &  1.2507 &  1.5759 &  1.9593 &  2.4081 &  2.9303\\
0003 &  0.3897 &  0.5398 &  0.7359 &  0.9862 &  1.3079 &  1.7027 &  2.1899 & &  0.7392 &  0.9674 &  1.2400 &  1.5615 &  1.9404 &  2.3838 &  2.8994\\
$\downarrow$ & & & & & & & & & & & & & & & \\
0999 &  7.5E-4 &  6.6E-5 & -0.0019 & -0.0023 &  0.0014 & -0.0098 &  0.0026 & & -0.0028 & -0.0042 & -0.0060 & -0.0078 & -0.0094 & -0.0116 & -0.0114\\
\enddata
\tablenotetext{a}{1000 $\Delta v$ values given in units of $\delta v_\parallel\!\left(z\right) \equiv 0.25246\,z + 1.4731$ km s$^{-1}$}
\tablecomments{These are the hybrid correlation 
values for full-resolution. [The complete version of this table can be found in the electronic edition of \apj\ or upon request.]}
\end{deluxetable}


\begin{deluxetable}{crrrrrrrrrrrrrrr}
\tablecolumns{16}
\tabletypesize{\tiny}
\tablewidth{7.25in}
\tablecaption{Cross-Correlation (\citet{2005kir360mnras1373} Mean Flux Decrement)\label{tab_cross_k05}}
\tablehead{
\colhead{$\Delta v$\tablenotemark{a}} & \multicolumn{7}{c}{$10\times\xi_\perp\!\left(\Delta v\right)$ in real-space} & 
\colhead{} & \multicolumn{7}{c}{$10\times\xi_\perp\!\left(\Delta v\right)$ in \emph{z}-space}\\
\cline{2-8}
\cline{10-16}
\colhead{($\delta v_\perp\!\left(z\right)$)} & 
\colhead{$z=1.8$} & \colhead{2.0} & \colhead{2.2} & \colhead{2.4} & 
\colhead{2.6} & \colhead{2.8} & \colhead{3.0} & 
\colhead{} & 
\colhead{$z=1.8$} & \colhead{2.0} & \colhead{2.2} & \colhead{2.4} & 
\colhead{2.6} & \colhead{2.8} & \colhead{3.0}}
\startdata
0000 &  0.3219 &  0.4503 &  0.5798 &  0.7403 &  0.9426 &  1.1868 &  1.4890 & &  0.5778 &  0.7721 &  0.9534 &  1.1649 &  1.4131 &  1.7022 &  2.0396\\
0001 &  0.3173 &  0.4449 &  0.5735 &  0.7323 &  0.9326 &  1.1738 &  1.4715 & &  0.5596 &  0.7527 &  0.9342 &  1.1448 &  1.3905 &  1.6772 &  2.0107\\
0002 &  0.3078 &  0.4319 &  0.5569 &  0.7110 &  0.9052 &  1.1389 &  1.4279 & &  0.5400 &  0.7295 &  0.9081 &  1.1144 &  1.3555 &  1.6371 &  1.9642\\
0003 &  0.2950 &  0.4143 &  0.5340 &  0.6817 &  0.8678 &  1.0918 &  1.3684 & &  0.5204 &  0.7049 &  0.8793 &  1.0807 &  1.3151 &  1.5886 &  1.9065\\
$\downarrow$ & & & & & & & & & & & & & & & \\
0300 &  0.0057 &  0.0074 &  0.0119 &  0.0148 &  0.0139 &  0.0119 &  0.0274 & &  0.0380 &  0.0400 &  0.0443 &  0.0570 &  0.0707 &  0.0786 &  0.0870\\
\enddata
\tablenotetext{a}{74 $\Delta v$ values given in units of $\delta v_\perp\!\left(z\right) \equiv 0.67455\,z + 0.38896$ km s$^{-1}$ (or, alternatively, arcseconds)}
\tablecomments{These are the hybrid correlation 
values for full-resolution. [The complete version of this table can be found in the electronic edition of \apj\ or upon request.]}
\end{deluxetable}


\begin{deluxetable}{crrrrrrrrrrrrrrr}
\tablecolumns{16}
\tabletypesize{\tiny}
\tablewidth{7.25in}
\tablecaption{Autocorrelation (\citet{2005kir360mnras1373} Mean Flux Decrement)\label{tab_auto_k05}}
\tablehead{
\colhead{$\Delta v$\tablenotemark{a}} & \multicolumn{7}{c}{$10\times\xi_\parallel\!\left(\Delta v\right)$ in real-space} & 
\colhead{} & \multicolumn{7}{c}{$10\times\xi_\parallel\!\left(\Delta v\right)$ in \emph{z}-space}\\
\cline{2-8}
\cline{10-16}
\colhead{($\delta v_\parallel\!\left(z\right)$)} & 
\colhead{$z=1.8$} & \colhead{2.0} & \colhead{2.2} & \colhead{2.4} & 
\colhead{2.6} & \colhead{2.8} & \colhead{3.0} & 
\colhead{} & 
\colhead{$z=1.8$} & \colhead{2.0} & \colhead{2.2} & \colhead{2.4} & 
\colhead{2.6} & \colhead{2.8} & \colhead{3.0}}
\startdata
0000 &  0.3219 &  0.4503 &  0.5798 &  0.7403 &  0.9426 &  1.1868 &  1.4890 & &  0.5778 &  0.7721 &  0.9534 &  1.1649 &  1.4131 &  1.7022 &  2.0396\\
0001 &  0.3162 &  0.4435 &  0.5721 &  0.7312 &  0.9320 &  1.1743 &  1.4742 & &  0.5770 &  0.7709 &  0.9518 &  1.1628 &  1.4103 &  1.6987 &  2.0352\\
0002 &  0.3033 &  0.4275 &  0.5531 &  0.7087 &  0.9052 &  1.1425 &  1.4362 & &  0.5746 &  0.7674 &  0.9471 &  1.1565 &  1.4022 &  1.6883 &  2.0221\\
0003 &  0.2868 &  0.4065 &  0.5277 &  0.6782 &  0.8685 &  1.0985 &  1.3834 & &  0.5707 &  0.7616 &  0.9393 &  1.1463 &  1.3890 &  1.6715 &  2.0007\\
$\downarrow$ & & & & & & & & & & & & & & & \\
0999 &  6.4E-4 &  3.2E-5 & -0.0011 & -0.0015 &  0.0011 & -0.0047 &  0.0012 & & -0.0021 & -0.0030 & -0.0041 & -0.0054 & -0.0067 & -0.0083 & -0.0084\\
\enddata
\tablenotetext{a}{1000 $\Delta v$ values given in units of $\delta v_\parallel\!\left(z\right) \equiv 0.25246\,z + 1.4731$ km s$^{-1}$}
\tablecomments{These are the hybrid correlation 
values for full-resolution. [The complete version of this table can be found in the electronic edition of \apj\ or upon request.]}
\end{deluxetable}

\clearpage

\appendix

\section{Autocorrelation Calculation for Arbitrary Spectral Resolution}\label{proof}

One way to compute the simulated correlation function corresponding to a given
spectral resolution is to carry out the calculations using spectra which have been individually
smoothed as appropriate.  However, assuming a Gaussian line spread function
(LSF), the autocorrelation 
curve for data of arbitrary spectral resolution can also be obtained by simply convolving
the full resolution curve with the LSF broadened
by a factor of $\sqrt{2}$ (eqs.~\ref{eq_auto},
\ref{eq_smoothing}, \ref{eq_autosigma1}, and \ref{eq_autosigma2}).  
The validity of this relation for our discrete, periodic simulated
spectra has been tested and verified.  Here, in the interest of clarity, we demonstrate
its origin for the simplified case of continuous spectra.  In this limit, equations~\ref{eq_auto},
\ref{eq_smoothing}, and \ref{eq_autosigma1} become

\begin{equation}\label{eq_auto_cont}
  \hat{\xi}_{\parallel}(\Delta v) = 
  \frac{1}{N_s} \sum_{n=1}^{N_s} 
  \frac{1}{N_\perp} \sum_{i=1}^{N_\perp}
  \frac{1}{\int dv_\parallel} \int
  \hat{\delta}_n\left(v_{\parallel}, v_{\perp_i}\right) \,
  \hat{\delta}_n\left(v_{\parallel}+\Delta v, v_{\perp_i}\right)
  dv_\parallel,
\end{equation}

\begin{equation}\label{eq_smoothing_cont}
  \hat{\mathcal{S}}_\parallel\left[\hat{f}\left(v_{\parallel}\right), \sigma\right] 
  = \int \hat{f}\left(\tau\right) \,
  \frac{1}{\sqrt{2\pi}\,\sigma} \,\,
  e^{\frac{-(v_{\parallel}-\tau)^2}{2\,\sigma^2}}
  d\tau,
\end{equation}

\noindent and

\begin{equation}\label{eq_autosigma1_cont} 
  \hat{\xi}^\sigma_\parallel\left(\Delta v\right) =
  \frac{1}{N_s} \sum^{N_s}_{n=1} 
  \frac{1}{N_\perp} \sum^{N_\perp}_{i=1} 
  \frac{1}{\int dv_\parallel} \int
  \hat{\mathcal{S}}_\parallel\left[\hat{\delta}_n\left(v_{\parallel}, v_{\perp_i}\right),
  \sigma\right] \,
  \hat{\mathcal{S}}_\parallel\left[\hat{\delta}_n\left(v_{\parallel}+\Delta v, v_{\perp_i}\right),
  \sigma\right]
  dv_\parallel,
\end{equation}

\noindent respectively.  Combining (\ref{eq_smoothing_cont}) and
(\ref{eq_autosigma1_cont}), and rearranging, yields

\begin{equation} 
  \hat{\xi}^\sigma_\parallel\left(\Delta v\right) =
  \frac{1}{N_s} \sum^{N_s}_{n=1} 
  \frac{1}{N_\perp} \sum^{N_\perp}_{i=1} 
  \frac{1}{\int dv_\parallel} \int \int 
  \hat{\delta_n}\left(\tau, v_{\perp i}\right) \,
  \hat{\delta_n}\left(\tilde{\tau}, v_{\perp i}\right) \, 
  \int 
  \frac{1}{2\pi\sigma^2} \,
  e^{\frac{-(v_{\parallel}-\tau)^2}{2\,\sigma^2}}
  e^{\frac{-(v_{\parallel}+ \Delta v -\tilde{\tau})^2}{2\,\sigma^2}}
  dv_\parallel \, d\tau \, d\tilde{\tau}.
\end{equation}

\noindent After substituting $\beta \equiv v_\parallel - \tau$ and noting that
$\int \frac{1}{2 \pi \sigma^2} \, e^{\frac{-\beta^2}{2 \sigma^2}}
e^{\frac{-(\beta-\gamma)^2}{2 \sigma^2}} d\beta = \frac{1}{2
  \sqrt{\pi} \sigma} e^{\frac{-\gamma^2}{4 \sigma^2}}$, this becomes

\begin{equation} 
  \hat{\xi}^\sigma_\parallel\left(\Delta v\right) =
  \frac{1}{N_s} \sum^{N_s}_{n=1} 
  \frac{1}{N_\perp} \sum^{N_\perp}_{i=1} 
  \frac{1}{\int dv_\parallel} \int \int 
  \hat{\delta_n}\left(\tau, v_{\perp i}\right) \,
  \hat{\delta_n}\left(\tilde{\tau}, v_{\perp i}\right) \, 
  \frac{1}{2 \sqrt{\pi} \sigma} \,
  e^{\frac{-(\tilde{\tau} - \tau - \Delta v)^2}{4 \sigma^2}}
  d\tau \, d\tilde{\tau}.
\end{equation}

\noindent With another change of variables ($\tilde{\tau} \equiv \tau
+ \alpha$), $\hat{\xi}^\sigma_\parallel$ can now be written in terms of
(\ref{eq_auto_cont}) and (\ref{eq_smoothing_cont}):

\begin{eqnarray}
  \hat{\xi}^\sigma_\parallel\left(\Delta v\right) & = &
  \frac{1}{N_s} \sum^{N_s}_{n=1} 
  \frac{1}{N_\perp} \sum^{N_\perp}_{i=1} 
  \frac{1}{\int dv_\parallel} \int \int 
  \hat{\delta_n}\left(\tau, v_{\perp i}\right) \,
  \hat{\delta_n}\left(\tau+\alpha, v_{\perp i}\right) \, 
  \frac{1}{2 \sqrt{\pi} \sigma} \,
  e^{\frac{-(\alpha - \Delta v)^2}{4 \sigma^2}}
  d\tau \, d\alpha
\\
  \hat{\xi}^\sigma_\parallel\left(\Delta v\right) & = &
  \int \left(
  \frac{1}{N_s} \sum^{N_s}_{n=1} 
  \frac{1}{N_\perp} \sum^{N_\perp}_{i=1} 
  \frac{1}{\int dv_\parallel} \int 
  \hat{\delta_n}\left(\tau, v_{\perp i}\right) \,
  \hat{\delta_n}\left(\tau+\alpha, v_{\perp i}\right) \, 
  d\tau \right)
  \frac{1}{\sqrt{2\pi} (\sqrt{2}\sigma)} \,
  e^{\frac{-(\alpha - \Delta v)^2}{2 (\sqrt{2}\sigma)^2}}
  d\alpha
\\
  \hat{\xi}^\sigma_\parallel\left(\Delta v\right) & = &
  \hat{\mathcal{S}}_\parallel\left[\hat{\xi}_\parallel\left(\Delta
  v\right), \sqrt{2}\sigma\right].
\end{eqnarray}

\clearpage 

\end{document}